\documentclass[numberedappendix]{emulateapj}
\usepackage{graphicx}
\usepackage{amssymb}
\usepackage{amsmath}
\usepackage{esdiff}
\usepackage{multirow}
\usepackage{chngcntr}
\usepackage{hyperref}

\newcommand{\ha}{\ensuremath{\rm H \alpha}}
\newcommand{\hi}{HI}
\newcommand{\hii}{HII}
\newcommand{\m}{$\mu$m}

\def\avg#1{\langle #1 \rangle}

\begin{document}

\title{Spatial and Spectral Modeling of the Gamma-ray Distribution in the Large Magellanic Cloud}
\author{Gary Foreman \altaffilmark{1}, 
You-Hua Chu \altaffilmark{1}, 
Robert Gruendl \altaffilmark{1}, 
Annie Hughes \altaffilmark{2},
Brian Fields \altaffilmark{1,3}, and 
Paul Ricker \altaffilmark{1}}
\altaffiltext{1}{Department of Astronomy, University of Illinois, 1002 W.\ Green St., Urbana, IL 61801, USA; \email{gforema2@illinois.edu}}
\altaffiltext{2}{Max-Planck-Institut f\"{u}r Astronomie, K\"{o}nigstuhl 17, D-69117 Heidelberg, Germany}
\altaffiltext{3}{Department of Physics, University of Illinois, 1110 W.\ Green St., Urbana, IL 61801, USA}

\begin{abstract}
We perform spatial and spectral analyses of the LMC gamma-ray emission collected over 66 months by the {\it Fermi Gamma-ray Space Telescope}. In our spatial analysis, we model the LMC cosmic-ray distribution and gamma-ray production using observed maps of the LMC interstellar medium, star formation history, interstellar radiation field, and synchrotron emission. We use bootstrapping of the data to quantify the robustness of spatial model performance. We model the LMC gamma-ray spectrum using fitting functions derived from the physics of $\pi^0$ decay, bremsstrahlung, and inverse Compton scattering. We find the integrated gamma-ray flux of the LMC from 200 MeV to 20 GeV to be $1.37 \pm 0.02 \times 10^{-7}$ ph cm$^{-2}$ s$^{-1}$, of which we attribute about 6\% to inverse Compton scattering and 44\% to bremsstrahlung. From our work, we conclude that the spectral index of the LMC cosmic-ray proton population is 2.4$\pm$0.2, and we find that cosmic-ray energy loss through gamma-ray production is concentrated within a few 100 pc of acceleration sites. Assuming cosmic-ray energy equipartition with magnetic fields, we estimate LMC cosmic rays encounter an average magnetic field strength $\sim$3 $\mu$G. 
\end{abstract}

\section{INTRODUCTION}
The LMC is by far the brightest source 
of diffuse gamma-ray flux detected outside of the Milky Way. The LMC 
was first observed in gamma rays by the Energetic Gamma Ray Experiment Telescope
with significance $>4.5\sigma$ \citep{1992ApJ...400L..67S}. However, a spatial
comparison of the LMC gamma-ray emission with observations at other wavelengths
was not possible before the \textit{Fermi Gamma-ray Space Telescope}. The first
such spatial analysis was performed by \citet{2010A&A...512A...7A} who determined
\textit{Fermi} had detected the LMC at $33\sigma$ significance after 11 months
of observation. As an external, star-forming galaxy detected at high significance in gamma rays,
the LMC provides a cosmic-ray laboratory that complements the Milky Way, which has been extensively
modeled by the Galactic cosmic-ray propagation team \citetext{GALPROP, \citealp{1998ApJ...509..212S}}.

The LMC is at a well-constrained 
distance of 50 kpc \citep{2009MNRAS.397..933M, 2013Natur.495...76P} with a
modest inclination of approximately $30^{\circ}$ \citep{1972VA.....14..163D, 
1998ApJ...503..674K, 2001AJ....122.1807V}. This small inclination angle means
self extinction and source confusion are low in the LMC, and the galaxy's high
Galactic latitude means the LMC does not suffer greatly from extinction by
the Milky Way. If unaccounted for, the LMC's inclination angle introduces a
$<$10\% error on distance measurements, much less than distance uncertainties
to some objects in the Milky Way, e.g., SNR RCW 86 for which various studies
have adopted a distance anywhere from 2 to 3 kpc \citep{2014A&A...562A.141M}.
Furthermore, the LMC is near
enough that stars can be resolved and the interstellar medium
(ISM) can be studied with high resolution (a few pc), yet it 
is distant enough to allow a global view of its stars and ISM.
The detailed and global knowledge of stars and ISM empirically
anchor the analysis of gamma-ray observations and study of 
cosmic rays in the LMC.

The LMC hosts active star formation, creating massive stars that live 
and explode as supernovae in a few to a few tens of Myr.  
The massive stars' fast winds and final supernova explosions 
interact with the ambient ISM and generate shocks, which produce cosmic 
rays by diffusive shock acceleration
\citep{1949PhRv...75.1169F, 2011Sci...334.1103A, 2013Sci...339..807A}.

\citet{2001ApJ...558...63P} predicted that the LMC should be the
brightest external normal star-forming galaxy seen by {\it Fermi},
along with the SMC.
{\it Fermi} has confirmed these predictions and firmly established
star-forming galaxies as a gamma-ray source class, which has lead to
recent theoretical studies such as \citet{2012JPhCS.355a2038P} and
\citet{2014A&A...564A..61M}. The SMC 
\citep{2010A&A...523A..46A}, M31 \citep{2010A&A...523L...2A}, M82 and 
NGC\,253 \citep{2010ApJ...709L.152A}, and two other starburst galaxies
are detected, but all of these sources do not have the spatial extent of
the LMC and are detected at a much lower signal-to-noise ratio in gamma rays.

The LMC stands out as the most intense, spatially extended source of 
gamma-ray emission outside the Milky Way; thus, the
\textit{Fermi} gamma-ray observations of the LMC render it an
invaluable laboratory for the study of cosmic-ray physics. Moreover,
star-forming galaxies are important contributors to the extragalactic 
gamma-ray background \citep{2002ApJ...575L...5P, 2010ApJ...722L.199F},
and so an understanding of the gamma-ray emission properties in individual
resolved galaxies is important for modeling the unresolved contribution to
the background.

Cosmic rays propagate through their host galaxy until they 
either escape or lose their energy through photon emission and 
particle-particle collisions. 
Diffuse gamma-ray emission is often produced along the way 
by cosmic-ray protons and electrons via three different
mechanisms.
Cosmic-ray protons lose energy via hadronic gamma-ray 
emission when they collide with thermal ISM protons to produce 
neutral pions that quickly decay to gamma rays: 
\begin{equation}\label{eq:pionInteraction}
p^+_{\rm CR} + p^+_{\rm ISM} \rightarrow 
p^+ + p^+ + \pi^0,\ \pi^0 \rightarrow \gamma + \gamma. 
\end{equation}
Cosmic-ray electrons emit bremsstrahlung radiation 
at gamma-ray energies when they scatter off of interstellar
nuclei:
\begin{equation}\label{eq:bremInteraction}
e^-_{\rm CR} + p^+_{\rm ISM} \rightarrow
e^- + p^+ + \gamma
\end{equation}
Cosmic-ray electrons can also up-scatter interstellar radiation 
photons to gamma-ray energies via inverse Compton scattering: 
\begin{equation}\label{eq:ICInteraction}
e^-_{\rm CR} + \gamma_{\rm ISRF} \rightarrow e^- 
+ \gamma_{\rm \gamma-ray}.
\end{equation} 
The spatial analysis performed by \citet{2010A&A...512A...7A} indicates the
{\it Fermi} detection of the LMC is likely diffuse, as expected if these interactions
are indeed the sources of the gamma-ray emission. The highly significant, spatially extended
{\it Fermi} detection of the LMC, complemented 
by detailed and global knowledge of the distribution of stars and ISM, allows us
to empirically test the relative importance of these
three mechanisms in generating gamma rays.

Here we investigate the relative contributions to the LMC gamma-ray signal
by each of the cosmic-ray interactions given in Eqs.~(\ref{eq:pionInteraction})--(\ref{eq:ICInteraction}).
We perform a binned
likelihood analysis to determine where in the LMC these interactions occur. We
expand on the spatial analysis of \citet{2010A&A...512A...7A} by:
\begin{enumerate}
\item Using the distribution of LMC star formation and tracers of the LMC's interstellar radiation
field (ISRF) as quantitative probes of gamma-ray intensity. 
\item Enforcing stricter photon selection criteria for better angular resolution given the luxury of more data.
\item Bootstrapping the {\it Fermi} data to test robustness of spatial model rankings.
\item Applying, across the entire galaxy, the smoothing kernels used 
\citet{2012ApJ...750..126M} in their cosmic-ray diffusion analysis of
the LMC star-forming region 30 Doradus (30 Dor).
\end{enumerate}

In an independent analysis, we assess the spatially integrated 
LMC gamma-ray spectrum as a means for predicting the flux contributions 
of the gamma-ray producing cosmic-ray interactions. We simultaneously
fit three theoretical spectra (pion decay, bremsstrahlung, and inverse Compton)
to the LMC spectrum. The fitted spectra are based on the
one-zone model presented by \citet{2013ApJ...773..104C} in their study of
inverse Compton gamma-ray emission in star-forming galaxies where we have
rescaled those authors' model of the Milky Way to appropriately model the LMC. Our method
expands on the work of \citet{2010A&A...512A...7A} by independently fitting
each of these three spectra rather than assuming their relative ratios
and fitting a single spectral model of their sum.

This article proceeds as follows: \S\ref{sec:Data} describes 
our data selection and preparation. In \S\ref{sec:BG}, we 
list our sources of background contamination and the means 
by which we remove them. \S\ref{sec:TimeVary} addresses the 
possibility that a point source unassociated with the diffuse 
emission of the LMC may be in or behind the galaxy. 
In \S\ref{sec:Models}, we motivate our models for the LMC 
gamma rays. In \S\ref{sec:Results} and \S\ref{sec:Spectral}, we present the results 
of our spatial and spectral analyses respectively. We discuss our results in the context of 
past studies and motivate future work in \S\ref{sec:Discussion},
and we make our concluding remarks in \S\ref{sec:Conclusions}.

\section{DATA}\label{sec:Data}
\begin{figure}[ht]
\begin{center}
\includegraphics[width=3.0in]{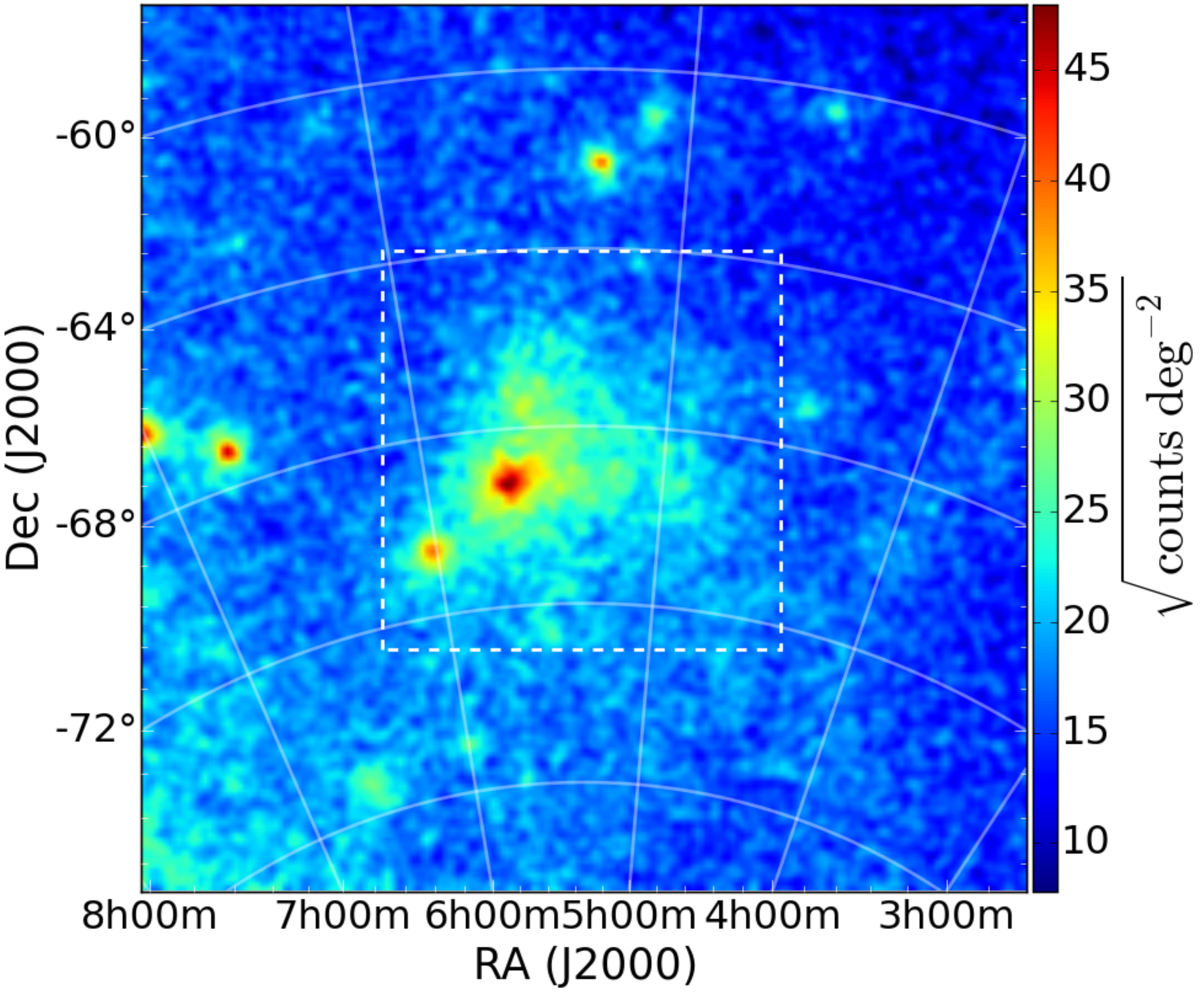}
\figcaption{\textit{Fermi} data from a $20^{\circ} \times 20^{\circ}$ region surrounding the LMC. Photons range in energy from 200 MeV to 20 GeV and were collected over a period of about 66 months. The dashed box shows the extent of the images in Figure~\ref{fig:hadron}. A square root color scale is used for display purposes only, and the image has been convolved with a $\sigma = 0.1^\circ$ Gaussian smoothing kernel. \label{fig:binned_counts}}
\end{center}
\end{figure}

Following \citet{2010A&A...512A...7A}, we select a 20$^{\circ}$ $\times$ 20$^{\circ}$ region surrounding the LMC (centered at 5$^{\mathrm{h}} 20^{\mathrm{m}}$, $-68^{\circ} 30'$). This large region of interest that extends well beyond the LMC gamma-ray emission is crucial for modeling of background sources (see \S\ref{sec:BG}). These data were collected by the \textit{Fermi} Large Area Telescope from 2008 August 8 to 2014 January 30, and the gamma rays we analyze range in energy from 200 MeV to 20 GeV. Figure~\ref{fig:binned_counts} shows the raw data with a \texttt{TAN} projection \citep{2002A&A...395.1077C}. We chose this projection of the celestial sphere, which differs from that used in \citet{2010A&A...512A...7A}, for easier comparisons with images at other wavelengths, e.g.\ Figures~\ref{fig:hadron} and \ref{fig:ISRF}.

We perform a binned maximum likelihood analysis\footnote{See \url{http://fermi.gsfc.nasa.gov/ssc/data/analysis/scitools/binned_likelihood_tutorial.html} for details.} of the data described above using the following method. We bin the gamma-ray photons detected by \textit{Fermi} into $0^{\circ}.1 \times 0.1^{\circ}$ cells, thereby creating a $200 \times 200$ pixel map of the region of interest. Front-converting photons of energies between 200 MeV and 20 GeV are placed into six logarithmically spaced energy bins for fitting of a power-law spectrum. By selecting only front converting photons, we are able to enhance the angular resolution of our data. We count photons only when the LMC is within 100$^{\circ}$ of zenith so that gamma rays originating from Earth do not contaminate the data. Finally, we adopt the \texttt{P7REP\_CLEAN\_V15} instrument response function as recommended by the \textit{Fermi} Science Support Center.

Our analysis of the gamma rays from the LMC has used numerous maps of the stars and ISM. These maps, introduced below, are shown in arbitrary units because the normalizations are ultimately left as free parameters in our binned maximum likelihood analysis. Our aim here is to familiarize the reader with the various spatial distributions of the LMC's interstellar components. We refer the reader to the references given below for further details about each map, and we give a more detailed description of the physics these maps represent in \S~\ref{sec:Models}. The gamma-ray map shown in Figure~\ref{fig:binned_counts} has coarse spatial binning compared with measurements at other wavelengths. We regrid the maps presented below to match the $200 \times 200$ pixel TAN projection of Figure~\ref{fig:binned_counts}.

To model gas-dominated (hadronic + bremsstrahlung) gamma-ray emission, we use the maps shown in Figure~\ref{fig:hadron}. The top image in Figure~\ref{fig:hadron} is a 656.3 nm map of the LMC from the Southern H$\alpha$ Sky Survey Atlas (SHASSA) taken using the El Enano telescope at Cerro Tololo Inter-American Observatory \citep{2001PASP..113.1326G}. It has been extinction corrected using the 160 \m\ image of \citet{2006AJ....132.2268M} (see Appendix~\ref{app:nthRadio} for details). The middle image in Figure~\ref{fig:hadron} is an HI 21-cm line map of the LMC using combined data from the Australia Telescope Compact Array and the Parkes Telescope \citep{2003ApJS..148..473K}. Here, we convert intensity to column density by assuming the neutral hydrogen emission to be optically thin. Following \citet{2010A&A...512A...7A} and \citet{bernardetal08}, we assume any presence of a dark neutral medium is well traced by the HI map. The bottom image is a map of molecular hydrogen as estimated from the NANTEN CO survey data performed by \citet{2001PASJ...53L..41F}, where \citet{2007ApJ...671..374Y} have converted from CO to H$_2$ column using $X_{\rm CO} = 5.4 \times 10^{20}$ H$_2$ atoms cm$^{-2}$ (K km s$^{-1}$)$^{-1}$.

{\begin{figure}[!ht]
\centering
\includegraphics[width=2in]{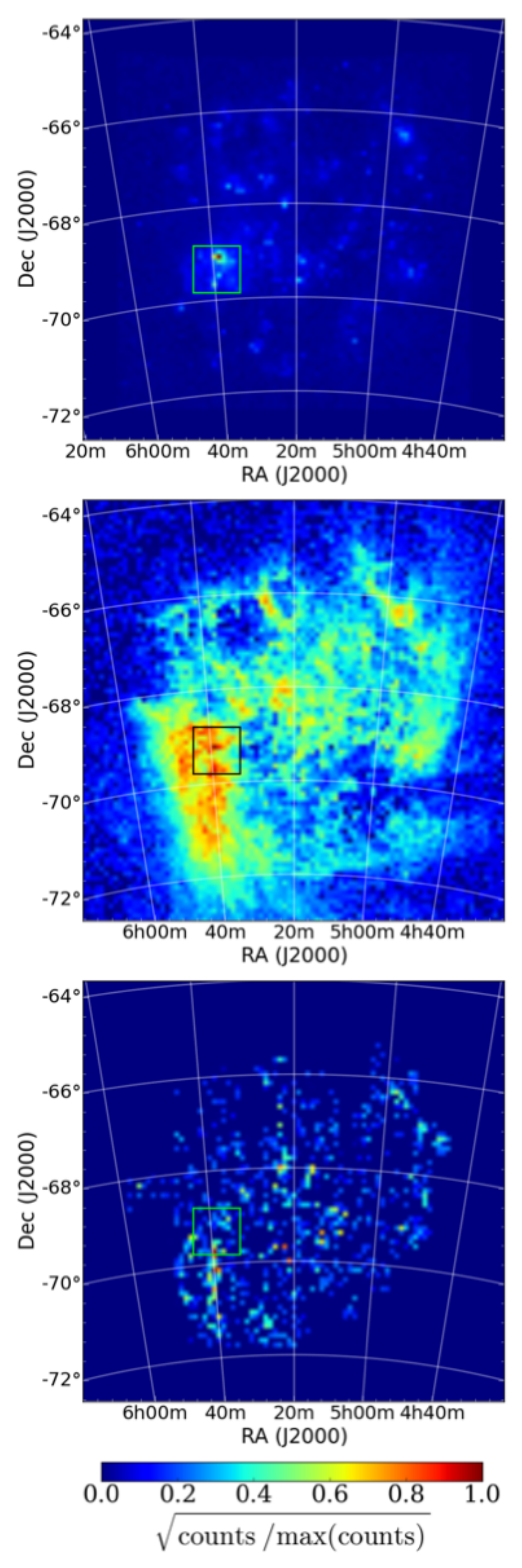}
\figcaption{$9^{\circ} \times 9^{\circ}$ ($90 \times 90$ pixel) images of LMC components used to model hadronic and bremsstrahlung gamma-ray emission. Images are normalized so that their maximum pixel values equal one, and the color scales are square root. The boxes indicate the star-forming region 30 Doradus. \textit{Top}: H$\alpha$ emission \citep{2001PASP..113.1326G}. \textit{Middle}: HI emission \citep{2003ApJS..148..473K}. \textit{Bottom}: H$_2$ column density estimated by \citet{2007ApJ...671..374Y} using CO data from \citet{2001PASJ...53L..41F}. \label{fig:hadron}}
\end{figure}}

We use the images presented in Figure~\ref{fig:ISRF} to represent the ISRF of the LMC. We selected these images based on the global spectral energy distribution (SED) of the LMC \citep{2010A&A...519A..67I, 2013AJ....146...62M, 2013arXiv1308.4284K}. The SED peaks in the near infrared at $\sim$1 \m\ and in the far infrared at $\sim$200 \m. The top image of Figure~\ref{fig:ISRF} shows the distribution of light from low-mass stars at 1.24~\m. These data are part of the 2MASS \citep{2006AJ....131.1163S}. The bottom image of Figure~\ref{fig:ISRF} traces infrared-emitting dust at 160 \m. This map was obtained using the MIPS instrument on the \textit{Spitzer Space Telescope} as part of the Surveying the Agents of a Galaxy's Evolution (SAGE) program \citep{2006AJ....132.2268M}.

{\begin{figure}[!ht]
\centering
\includegraphics[width=2in]{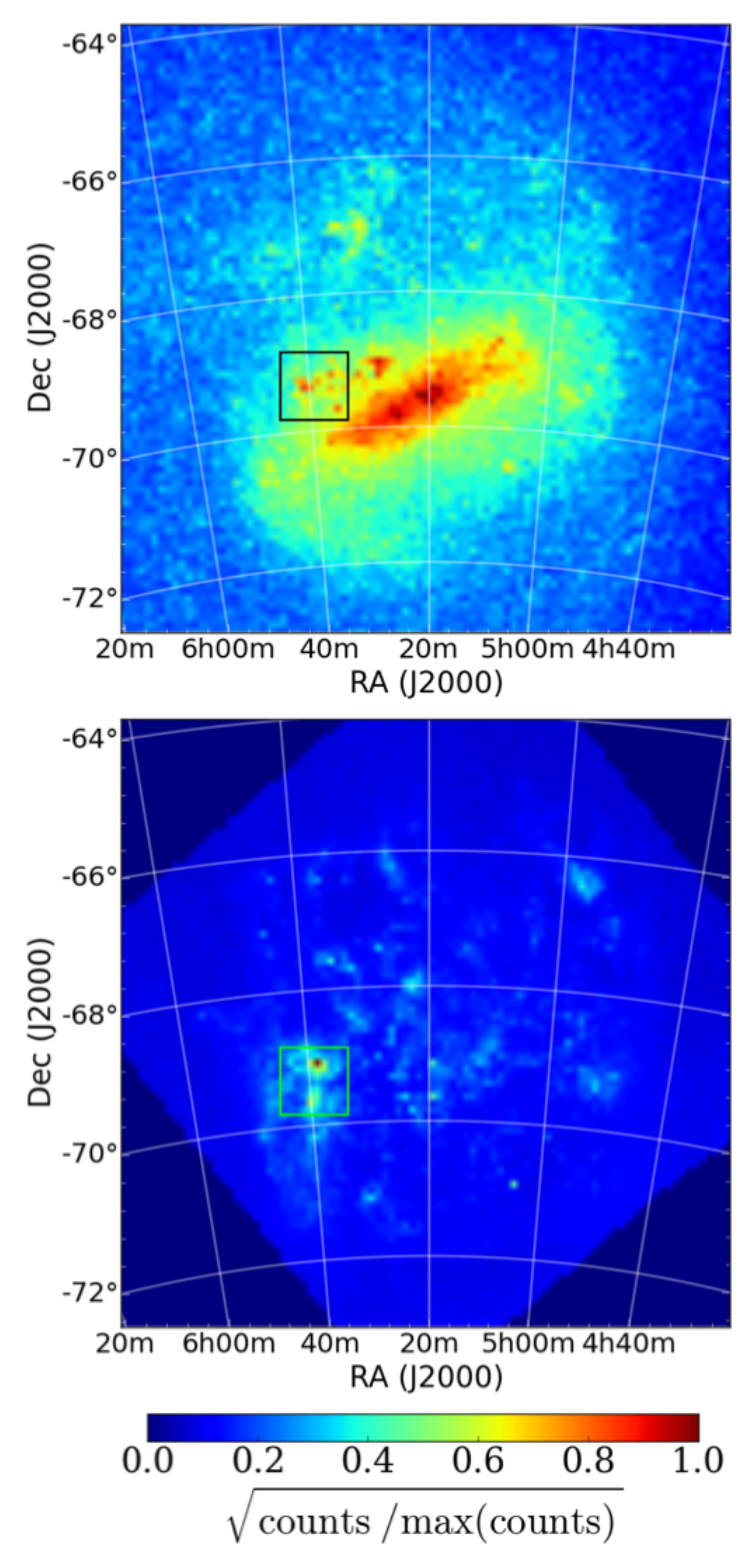}
\figcaption{$9^{\circ} \times 9^{\circ}$ ($90 \times 90$ pixel) images of LMC components used to model the interstellar radiation field. The boxes indicate the star-forming region 30 Doradus. \textit{Top}: 2MASS 1.24 \m\ J band image \citep{2006AJ....131.1163S}. \textit{Bottom}: MIPS 160 \m\ image \citep{2006AJ....132.2268M}. \label{fig:ISRF}}
\end{figure}}

The data described above represent the distributions of targets for the collisional processes that ultimately lead to gamma-ray production, i.e., the ISM protons and ISRF photons with which cosmic-ray protons and electrons collide. Figure~\ref{fig:HZSF} shows one of our cosmic-ray flux models: star formation rates at 6.3 Myr (top) and 12.5 Myr (bottom) from \citet{2009AJ....138.1243H}. The authors generated these maps by adopting an initial mass function and by fitting a set of isochrones to data they collected with the $U$, $B$, $V$, $I$ filtering system on the 1 m Swope Telescope at Las Campanas Observatory as part of the Magellanic Cloud Photometric Survey \citep{1997AJ....114.1002Z}.

{\begin{figure}[!ht]
\centering
\includegraphics[width=2in]{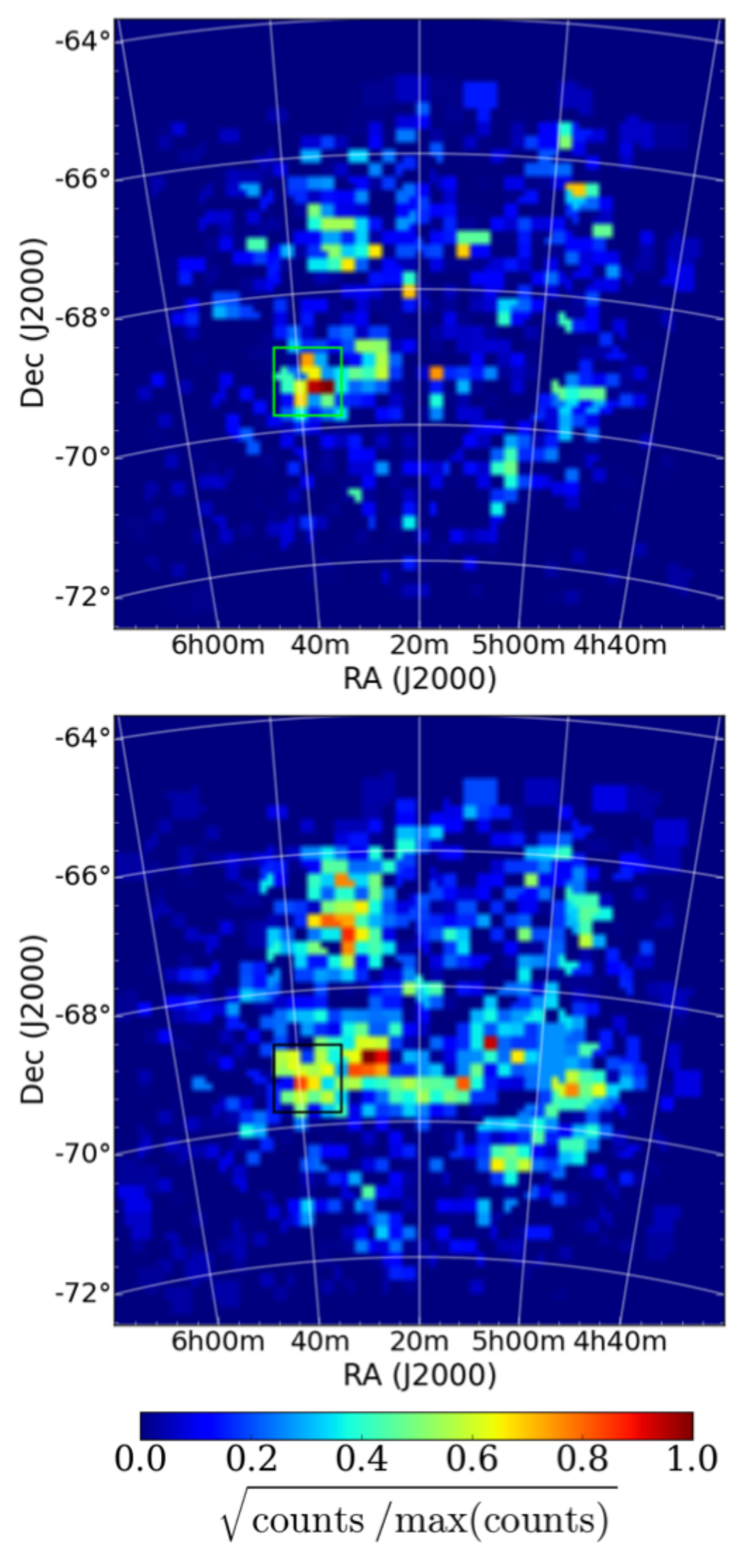}
\figcaption{$9^{\circ} \times 9^{\circ}$ ($90 \times 90$ pixel) star formation rate maps used for our concentrated (see \S~\ref{sec:Models}) cosmic-ray distribution models. \textit{Top}: 6.3 Myr star formation rate map \citep{2009AJ....138.1243H}. \textit{Bottom}: 12.5 Myr star formation rate map \citep{2009AJ....138.1243H}. \label{fig:HZSF}}
\end{figure}}

\begin{figure}[!ht]
\begin{centering}
\includegraphics[width=2in]{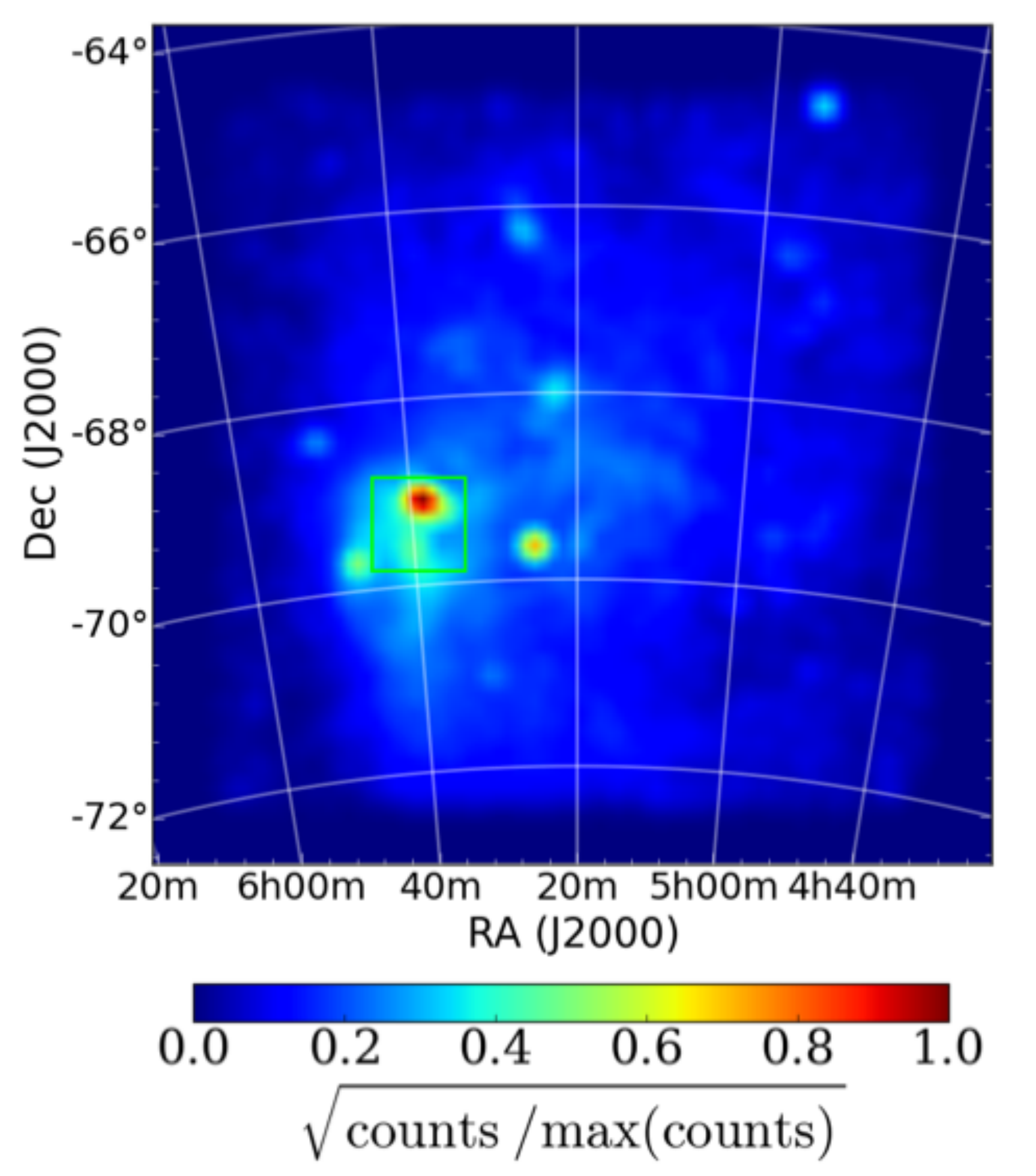}
\figcaption{Non-thermal 1.4 GHz map of the LMC used for our \textit{diffuse} cosmic-ray electron models. See \citet{2007MNRAS.382..543H} for details about the original radio data. For full resolution and information about the subtraction of the thermal component is given in Appendix~\ref{app:nthRadio} \label{fig:nthRadio}}
\end{centering}
\end{figure}

In Figure~\ref{fig:nthRadio}, we present a map of the non-thermal component of the 1.4 GHz radio continuum emission in the LMC, which traces the distribution of synchrotron-emitting cosmic-ray leptons. The original radio map used to generate this image was presented in \citet{2007MNRAS.382..543H} and is a combination of interferometric (ATCA) and single-dish (Parkes) data. We estimate and subtract the thermal contribution from the original radio data using the prescription of \citet{tabatabaeietal07}. We use a scaled version of the de-reddened H$\alpha$ map presented in Figure~\ref{fig:hadron} as an estimate for the thermal component. For full details about the construction of the map in Figure~\ref{fig:nthRadio} and for a full resolution image, see Appendix~\ref{app:nthRadio}.

\section{BACKGROUNDS AND FOREGROUNDS}\label{sec:BG}
The unprocessed data shown in Figure~\ref{fig:binned_counts} include not only diffuse emission from the LMC but also Galactic and extragalactic diffuse emission as well as background gamma-ray point sources. These contaminating sources must be removed in order to analyze the diffuse gamma-ray emission from the LMC. We use the \texttt{iso\_clean\_front\_v05.txt} model supplied with the \textit{Fermi} Science Tools package for the isotropic spectral template in our analysis. Similarly, \textit{Fermi} supplies the Milky Way model as \texttt{gll\_iem\_v05\_rev1.fits}. For each of these models, the likelihood estimator fits a multiplicative normalization factor that scales the spectrum defined by these two models. We also include the fourteen point sources in our region of interest reported by the \textit{Fermi} Large Area Telescope Second Source Catalog \citep{2012ApJS..199...31N}, as well as one SIMBAD blazar, PKS J0352-6831, which appears to have flared since the release of 2FGL. Most point sources are modeled using a power-law spectrum with two parameters: a normalization that scales the spectrum and a power law spectral index. The normalization factors of these point sources depend on their apparent brightnesses, and the spectral indices range from $-1.3$ to $-2.9$. We model the brightest point source, 2FGL J0601-7037, using a log parabola spectrum parameterized by a normalization and two indices describing the exponential energy dependence. The top image of Figure~\ref{fig:background} shows our  background model as fit by the binned likelihood analysis, and the bottom image shows the background subtracted 9$^{\circ}$ $\times$ 9$^{\circ}$ central region. Thin, solid contours indicate the quantity
\begin{equation}
\sigma_i = \frac{|n_i - m_i|}{\sqrt{m_i}},
\end{equation}
where $m_i$ is the number of counts predicted by the background model in the $i$th pixel, and $n_i$ is the number of counts observed in that pixel. These contours are placed at 1, 2, 3, 4, and 5$\sigma$. The 1$\sigma$ contour indicates the region used for our time variability analysis described below.

{\begin{figure}[!ht]
\centering
\includegraphics[width=2.8in]{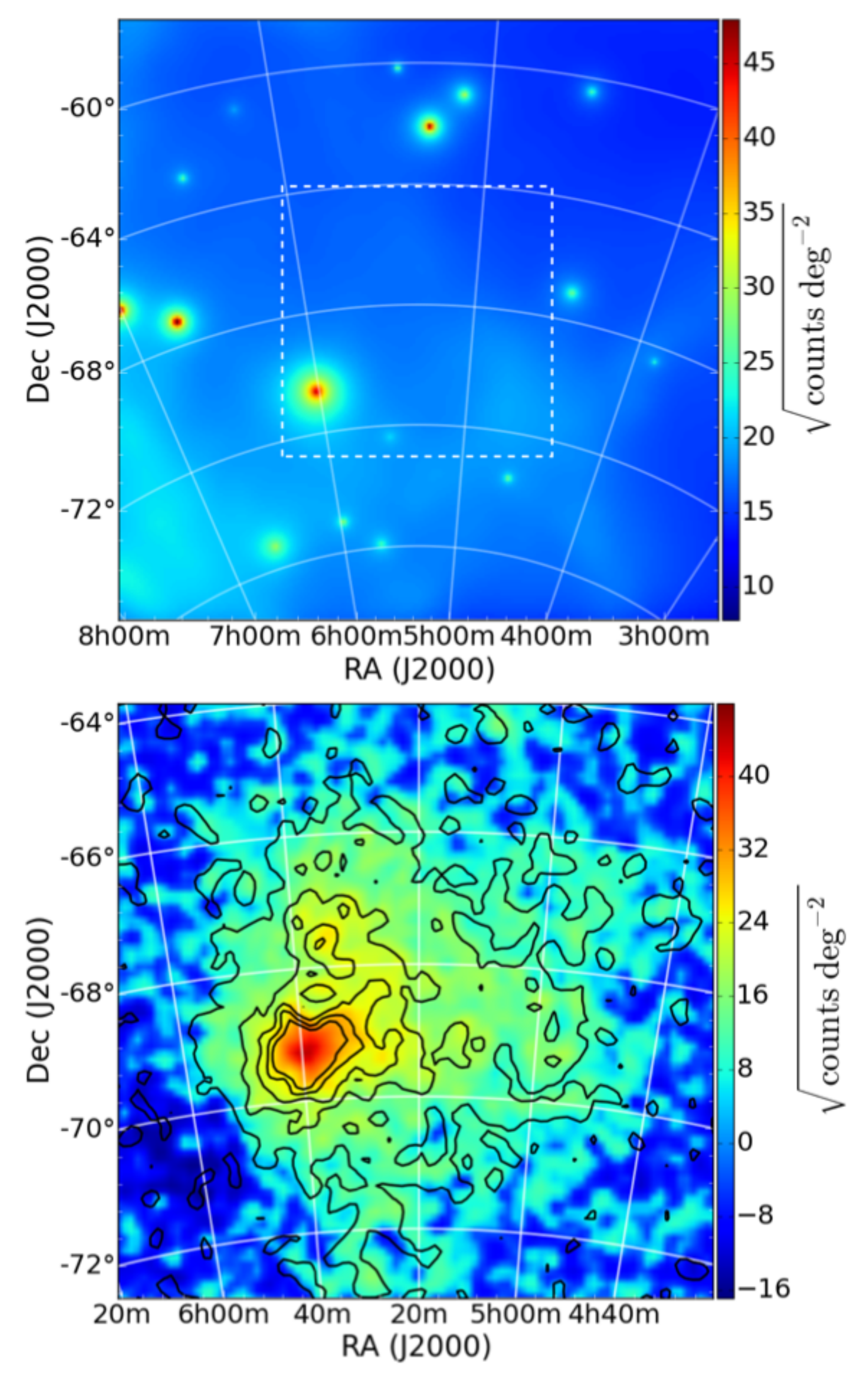}
\figcaption{\textit{Top}: LMC background model including fifteen point sources, Galactic foreground emission, and isotropic, diffuse emission. This model was fit to the data in Figure~\ref{fig:binned_counts} using a binned likelihood analysis with the \textit{Fermi} Science Tools software. Dashed box is the same as in Figure~\ref{fig:binned_counts}, and square root color scale is for display purposes only. \textit{Bottom}: background subtracted counts map. The color bar indicates the quantity sign(${\rm data - background}$)$\sqrt{|\mathrm{data} - \mathrm{background}|}$. The thin, solid contours show $\sigma = |\mathrm{data-background}| / \sqrt{\mathrm{background}}$ and are placed at 1, 2, 3, 4, and 5$\sigma$. The extent of this image is the same as the dashed box in the top image. \label{fig:background}}
\end{figure}}

\section{TIME VARIABILITY}\label{sec:TimeVary}
If the gamma-ray emission from the LMC is indeed caused by cosmic-ray energy losses, the gamma-ray flux should not vary appreciably on the time scales probed by \textit{Fermi}. Following \citet{2010A&A...512A...7A}, we perform a month-by-month analysis of our data sample to ensure the LMC data are not significantly contaminated, for example, by a background blazar.

In their analysis of the first year LMC data, \citet{2010A&A...512A...7A} found an unusually high gamma-ray flux for photons $>$100 MeV during the fourth month of observations and excluded this time period from their analysis. We also find enhanced flux in month four of our time series analysis of 200 MeV -- 20 GeV photons. Although our method of estimating flux differs from that of the {\it Fermi} study, we also agree that the enhanced emission originates in or near the 30 Doradus (30 Dor) star-forming region.

Because of the strict requirements we have used for event selection (front-converting, zenith angle $<$100$^{\circ}$, etc.), we have fewer photons per month in our data sample than did \citet{2010A&A...512A...7A}. To avoid dependency on a spatial and spectral model for estimating monthly LMC fluxes, we choose to perform photon counting within the 1$\sigma$ contour shown in Figure~\ref{fig:background}. Photon counts are converted to fluxes by dividing the monthly exposures calculated using the {\tt gtexpmap2} function of the {\it Fermi} Science Tools software. It is important to note that the fluxes estimated from this method are sums of the LMC, diffuse background, and Milky Way components. This does not affect our search for variability because, as for the LMC, these diffuse components will also not vary appreciably on the time scale of a month.

\begin{figure}[!ht]
\begin{center}
\includegraphics[width=3.0in]{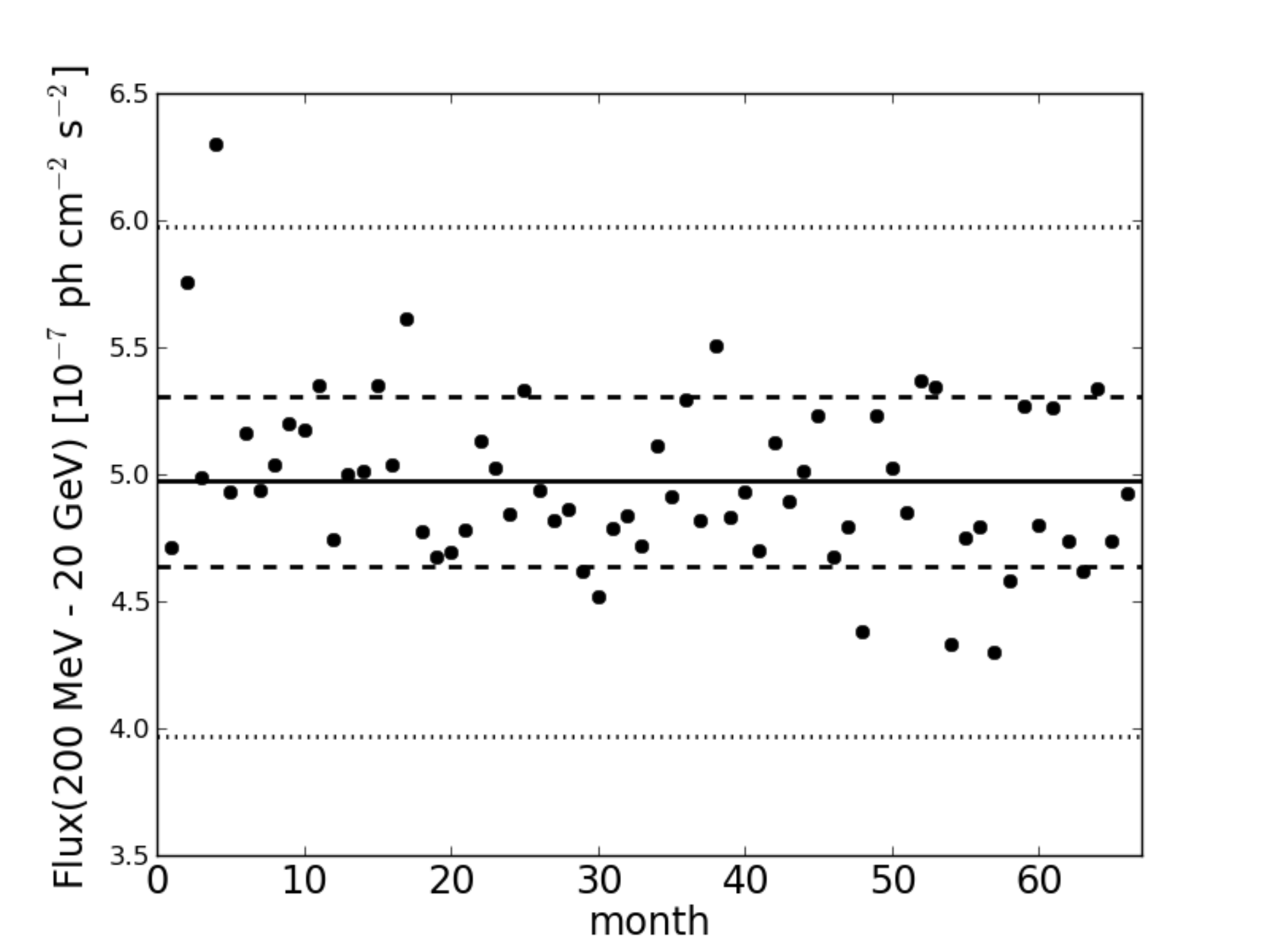}
\figcaption{LMC light curve determined by counting photons within the thick, dashed contour in Figure~\ref{fig:background}. The solid black line indicates the average monthly flux, the dashed lines indicate 1$\sigma$, and the dotted lines indicate 3$\sigma$. Month four is the only month with flux outside 3$\sigma$ from the mean. {\it Note}: these flux estimates include foreground and background photons and are, therefore, higher than fluxes reported from our likelihood analysis. \label{fig:lightcurve}}
\end{center}
\end{figure}

Figure~\ref{fig:lightcurve} shows the results of our time series analysis. Month four is the only month with flux that lies outside three standard deviations of the mean. Month four comprises $\sim$$1/66$ of our data, and we find that removing this month from our data would reduce the mean flux by 0.7\%. Compare this to the {\it Fermi} study in which month four accounted for $1/11$ of their data, and by our calculation, affected their flux estimate by 4\%. Given its $<1\%$ effect, and to avoid complicating the data reduction process, we choose to keep month four in our data set.

We repeat our time series analysis using the 2$\sigma$ contour shown in Figure~\ref{fig:background}, which encompasses the 30 Dor region and extends to the northwest. Again, the month four flux is a unique outlier greater than three standard deviations from the mean monthly flux. This supports the claim by \citet{2010A&A...512A...7A} that the enhanced emission originated in or near 30 Dor.

\section{SPATIAL ANALYSIS: MODELS}\label{sec:Models}
Our strategy is (i) to identify physically motivated models for the spatial character of the LMC gamma-ray emission, (ii) to connect these models to LMC observables at non-gamma-ray wavelengths, and (iii) to use spatial fits of the {\it Fermi} LMC signal to test these models. We perform a morphological analysis of the \textit{Fermi} gamma-ray map of the LMC using several combinations of the images in Figures~\ref{fig:hadron}, \ref{fig:ISRF}, \ref{fig:HZSF}, and \ref{fig:nthRadio}. These model images are then included in our binned likelihood analysis (described in \S~\ref{sec:Data}) as sources of type ``DiffuseSource'' and  spatial models of type ``SpatialMap.'' The normalization and the single-power-law spectral index are left as free parameters of each model map. Our models take the functional form
\begin{equation}
\label{eq:fitFunc}
I_{\gamma}(x, y, E) = \sum\limits_i a_i M_i(x, y) E^{-b_i},
\end{equation}
where $a_i$ and $b_i$ are the free parameters of the fit, $M_i$ is a combination of model maps described below, and $I_{\gamma}$ is the LMC gamma-ray intensity as a function of plane-of-sky position $(x, y)$ and energy $E$. The power-law dependence of intensity on energy well-describes the gamma-ray spectrum at $\gtrsim$1 GeV, but at lower energies, the spectrum turns over (see \S\ref{sec:Spectral}). The index given by Equation~(\ref{eq:fitFunc}) is an average over the entire energy range and will tend to underestimate the high energy power law behavior.

The models we choose from the maps in Figures~\ref{fig:hadron}, \ref{fig:ISRF}, \ref{fig:HZSF}, and \ref{fig:nthRadio} fall into three categories: gas models (\S~\ref{sec:hadron}), radiation models (\S~\ref{sec:lepton}), and a gas + radiation model (\S~\ref{sec:had+lep}). Additionally, we subclassify these models as diffuse and concentrated based on the quantity used to represent the cosmic-ray flux (further explanation below).

We perform our analysis with and without data from the highly active star-forming region 30 Doradus. The 30 Dor region dominates the gamma-ray, H$\alpha$, 6.3 Myr star formation, and synchrotron maps, and we remove it by subtracting all counts in the $1^{\circ} \times 1^{\circ}$ region from the model maps. We then model the 30 Dor gamma-rays separately as a point source. The region we remove is indicated by the boxes in Figures~\ref{fig:hadron}, \ref{fig:ISRF}, and \ref{fig:HZSF}.

\subsection{Gamma-ray Intensity}\label{sec:intensity}
The high-energy emission processes we
consider are from the interaction $i + j \rightarrow \gamma + \cdots$ 
of cosmic-ray particles
$i \in ({\rm e^-,\ ions})$ with interstellar target
$j \in (\rm gas,\ radiation)$.
The emissivity from such processes is schematically given by
\begin{equation}
q_\ell(E_\gamma)  = \frac{dN_\gamma}{dV dt dE_\gamma}= \langle \Phi_i \ n_j \sigma_{ij \rightarrow \gamma} \rangle ,
\end{equation}
where $\Phi_i$ is the flux of cosmic-ray species $i$,
and $n_j$ is the number density of target species $j$.
For a rigorous derivation of the gamma-ray emissivity,
see \citet{1971NASSP.249.....S}.
The cross section $\sigma$ for gamma-ray production depends on the
cosmic-ray energy and on the ambient photon energy in the case of 
inverse Compton. Thus, the emissivity averages over the appropriate
cosmic-ray spectrum and ISRF as indicated by angle brackets. Our 
spatial analysis does not integrate over energy dependence and 
instead uses observable proxies for cosmic-ray flux and target
densities.

Because the LMC is optically thin to gamma rays at {\em Fermi} energies
\citetext{e.g.\ \citealp{1971NASSP.249.....S}, \citealp{2006ApJ...640L.155M}},
the photon number intensity or surface brightness resulting from emission
process $\ell$ is just
\begin{equation}
I_\ell(E_\gamma) = \int_{\rm los} q_\ell(E_\gamma) \ dz,
\end{equation}
the line of sight integral of the emissivity, where we take the
$z$ axis to be along the sightline. 
Substituting our expression for emissivity, 
the intensity is a line integral of the product of cosmic-ray
flux and target density:
\begin{eqnarray}
\label{eq:intensity}
I_\ell &=& \int_{\rm los} \langle \Phi_i n_j \sigma_{ij \rightarrow \ell} \rangle \ dz \nonumber \\
 &=& \int_{\rm los} \langle \Phi_i \sigma_{ij \rightarrow \ell} \rangle \ dN_j
 \equiv \avg{\Phi_i \sigma_{ij \rightarrow \ell}} \ N_j \ .
\end{eqnarray}
We see that the intensity amounts to a weighted average
of cosmic-ray flux times target column density $N_j = \int n_j \ dz$.
Thus, we expect the gamma emissivity to trace the column density
of relevant targets.
In the case of gas targets, this is the column density of the 
gas phases in which the cosmic rays reside.
In the case of radiation targets (inverse Compton), the target
photon column density is, for an optically thin emission region, proportional
to the observed surface brightness.

From Equation~(\ref{eq:intensity}), we can define two limiting cases based on the cosmic-ray
flux distribution $\Phi_i$. In one limit, $\Phi_i$ does not vary spatially, which amounts
to assuming diffusion produces homogeneity in the cosmic rays and erases memory of
source positions. In this case, the target distributions uniquely trace the gamma-ray
distribution. In the other limit, diffusion has not yet occurred, and $\Phi_i$ is traced
by sites of cosmic-ray sourcing. We describe these limits in the context of our
spatial models below.

If the cosmic-ray spectra vary spatially---as is surely the
case at some level---then the gamma-ray spectrum will vary spatially
for each emission process. However, we will ignore this
effect both for simplicity purposes and because of the finite
amount of spatial information and spectral information in the {\em Fermi} data.
Previous studies of the LMC have made the same assumption \citep{2010A&A...512A...7A, 2012ApJ...750..126M}.

\subsection{Gas Models}\label{sec:hadron}
In the case of hadronic and bremsstrahlung gamma-ray emission, the target particles are interstellar gas atoms and molecules. The gamma-ray signal from these interactions traces the overlap of cosmic-ray ions and electrons with the LMC gas. In the limit where cosmic-ray flux is spatially uniform, we expect the hadronic and bremsstrahlung gamma-ray spatial signal to trace the total gas density. On the other hand, in the limit where cosmic rays are confined near their sources, we expect a gamma-ray signal concentrated near sites of recent star formation.

Figure~\ref{fig:hadron} shows maps of ionized, neutral, and molecular hydrogen. These three maps do not trace the thermal proton density in the same way. The HI and H$_2$ maps are proportional to column density, $N_{\mathrm{HI}} = \int^L_0 n_{\mathrm{H}} ds \approx n_{\mathrm{H}} L$, where $L$ is the path length along the line of site. The H$\alpha$ emission, on the other hand, results from recombination and is proportional to $\int^{L}_{0} n_p n_e ds$ $\approx$ $n_{\rm H}^2 L$, which is essentially density-weighted column density. Since $n_{\rm H}$ ranges from below 0.1 cm$^{-3}$ in the diffuse medium to over 10$^3$ cm$^{-3}$, using the H$\alpha$ surface brightness as a proxy for the ionized hydrogen column density can produce differential  errors by factors greater than 10$^4$.  We suggest the use of the square root of H$\alpha$ surface brightness as a proxy of ionized column density.  The square root of H$\alpha$ emission is proportional to $n_{\rm H} L^{1/2}$.  Because the emission path length ranges from 10 pc for individual HII region to 500 pc for the diffuse ionized gas (DIG) in the disk, the differential error introduced by our under-counting  the path length is less than a factor of 10. Thus, while not ideal, we consider the square root of H$\alpha$ as a far better proxy than H$\alpha$ for the ionized gas column density.

Our first gas models use the HI, H$_2$, \ha\ (included for comparison purposes) and $\sqrt{\mathrm{H}\alpha}$ maps by themselves. These models, the first three of which are used in the \citet{2010A&A...512A...7A} study, are akin to assuming an extreme case in which cosmic rays have diffusively propagated away from their acceleration sites in such a way as to create a uniform flux distribution across the LMC. With no spatial dependence, the isotropic cosmic-ray flux level then becomes one factor in the normalization parameter, $a_i$, in Equation~(\ref{eq:fitFunc}).

We explore the opposite extreme, a case in which cosmic rays remain in the immediate vicinity of their acceleration sites, by using the maps of star formation rates in Figure~\ref{fig:HZSF} as proxies for the cosmic-ray flux. These models are constructed such that the hydrogen maps (HI, H$_2$,  and $\sqrt{\mathrm{H}\alpha}$) are multiplied by each of the star formation rate maps, so that the binned likelihood software is given two separate functions with which to fit the LMC gamma rays. For example, our HI-times-star formation-rate model is formed by multiplying the HI map by both the 6.3 and 12.5 Myr star formation rate maps, i.e.,
\begin{equation}
M_i(x,y) = \left\{I_{\rm HI} \psi_{6.3 \rm Myr}, \hspace{3pt} I_{\rm HI} \psi_{12.5 \rm Myr}\right\},
\end{equation}
where $\psi_{6.3 \rm Myr}$ and $\psi_{12.5 \rm Myr}$ are the two epochs of star formation.
These two images are then simultaneously fit as separate components, each with its own normalization and spectral index. This method allows us to comment on the relative importance of the 6.3 and 12.5 Myr star formation epochs by comparing their relative flux contributions to each model. We combine the two maps fit on this first pass into one by adding them based on their flux contributions. After this step, the concentrated models take the form
\begin{equation}
M(x, y) = \frac{F_{6.3 \rm Myr}}{F_{\rm tot}} I_{\rm HI} \psi_{6.3 \rm Myr} + \frac{F_{12.5 \rm Myr}}{F_{\rm tot}} I_{\rm HI} \psi_{12.5 \rm Myr}.
\end{equation}
Finally, this single map is then refit to the data using a single normalization and a single spectral index pair. Maintaining the same number of free parameters for both diffuse and concentrated models eases their comparison using the likelihood statistic described below.

Conclusions about models that fit multiple images each with its own normalizations must be made cautiously; any physical offset between the images is washed out by the free renormalization performed by the likelihood estimator. For example, when \citet{2010A&A...512A...7A}, construct their $N({\rm HI}) + N({\rm H_2}) + N({\rm HII})$ model where they allow each hydrogen component to be fit independently, some information is lost. Specifically, the fact that HI typically has the greatest column density along a given site line is disregarded. Fortunately, the total LMC star formation rate is $\sim$0.4 M$_{\odot}$ yr$^{-1}$ during both the 6.3 and 12.5 Myr epochs \citep[Figure~11]{2009AJ....138.1243H}, so the two maps have virtually no physical offset.

We exclude from our discussion the older star formation rate maps of \citet{2009AJ....138.1243H} because when we included the 25 Myr map, it was not given an appreciable weighting by the likelihood analysis. This might suggest that cosmic rays accelerated during that epoch have undergone significant diffusion. The LMC was forming stars nearly 25\% more slowly 25 Myr ago \citep[Figure~11]{2009AJ....138.1243H}.

For completeness, we have also constructed a model of the form
\begin{equation}
M_i = I_{\rm HI} + I_{\rm H_2},
\end{equation}
as a ``total'' hydrogen-mass model (note this model is allowed only one normalization and one spectral index parameter). We have excluded H$\alpha$ from this model, first because $I_{\rm H\alpha}$ does not measure the same quantity as $I_{\rm HI}$ and $I_{\rm H_2}$ and also because ionized hydrogen does not contribute significantly to the total hydrogen mass \citep{2007ApJ...671..374Y}. 

\subsection{Radiation Models}\label{sec:lepton}
Inverse Compton emission is the result of scatterings between cosmic-ray electrons and photons of the ISRF. As with the gas models, we split the radiation models into diffuse and concentrated categories. Again, we use the star formation rate maps as a measure of the cosmic-ray flux in the concentrated models. We use the map presented in Figure~\ref{fig:nthRadio} for our diffuse models. The $\sim$3 GeV \citep{2012ApJ...750..126M} electrons emitting synchrotron radiation at 1.4 GHz are not themselves energetic enough to significantly contribute to the inverse Compton gamma-ray emission. However, this population of lower-energy electrons will be partially comprised of formerly high-energy electrons that have undergone inverse Compton scattering. These older, lower-energy electrons have had additional time to propagate after their scattering events and should be more diffuse.

Our choices in models for the ISRF are motivated by local maxima in the SEDs of \citet{2010A&A...519A..67I}, \citet{2013AJ....146...62M}, and \citet{2013arXiv1308.4284K}. We use the 1.24 \m\ map shown in Figure~\ref{fig:ISRF}, which traces the dominant energy contribution from stars, and we use the 160 \m\ map as a tracer for the energy contribution from dust. Although the cosmic microwave background (CMB) is not included in the SEDs mentioned above, it nevertheless contributes a relevant fraction of energy to the ISRF. The CMB is an isotropic source, so for our CMB models, in practice, we fit either the 1.4 GHz map or the star formation maps by themselves.

We note that the minimum electron energy required to scatter a 1.24 \m\ photon up to 200 MeV is given by
$E_e = \gamma_{\rm min} m_e c^2 = 0.5 \sqrt{\epsilon_{\gamma} / \epsilon_{1.24\ \mu{\rm m}}} m_e c^2 = 3.6\ {\rm GeV}.$
This is the lower limit energy for a cosmic-ray electron to contribute to the {\it Fermi} data we analyze here given the ISRF energies trace by the maps described above. So our previous statement that 1.4 GHz synchrotron-emitting cosmic-ray electrons do not contribute significantly to inverse Compton is well justified.

The map of synchrotron radiation traces the magnetic field strength in conjunction with the cosmic-ray electron density. We have made the simplifying assumption that at scales probed by {\it Fermi}, the magnetic field strength does not vary appreciably and is therefore a constant scale factor across the entire map.

\subsection{Gas + Radiation Model}\label{sec:had+lep}
We expect both the ISRF and the ISM to contribute as targets, and we explore models in which gas and radiation maps are fit simultaneously. These models take the form
\begin{eqnarray}
M_i(x,y) = \left\{I_{\rm gas} \left(\frac{F_{6.3 \rm Myr}}{F_{\rm tot, gas}} \psi_{6.3 \rm Myr} \right. \right. &+& \hspace{3pt} \left. \frac{F_{12.5 \rm Myr}}{F_{\rm tot, gas}} \psi_{12.5 \rm Myr}\right), \nonumber \\ 
I_{\rm rad} \left(\frac{F_{6.3 \rm Myr}}{F_{\rm tot, rad}} \psi_{6.3 \rm Myr} \right. &+& \hspace{3pt} \left. \left. \frac{F_{12.5 \rm Myr}}{F_{\rm tot, rad}} \psi_{12.5 \rm Myr} \right) \right\}. \nonumber \\
&&
\end{eqnarray}
The radiation and gas components are each fit using their own normalizations, $a_i$, and spectral indices, $b_i$, because we do not assume that inverse Compton and the collision processes involving gas targets result in similarly shaped gamma-ray spectra.

We do not test every combination of the gas and radiation maps described above because of resource limitations. We explain below which maps we have chosen for these combined models after discussing the performances of all single map models.

\section{SPATIAL ANALYSIS: RESULTS}\label{sec:Results}
As did \citet{2012ApJ...750....3A} in the {\it Fermi} group's study of the Milky Way gamma-ray emission, we use the likelihood statistic to distinguish between the spatial models described in the previous section. The likelihood as computed by the {\it Fermi} Science Tools software\footnote{See \url{http://http://fermi.gsfc.nasa.gov/ssc/data/analysis/documentation/Cicerone/Cicerone_Likelihood/Likelihood_formula.html} for details.} is given as
\begin{equation}
\mathrm{LH} = e^{-N_{\rm exp}} \prod_i \frac{m_i^{n_i}}{n_i!},
\end{equation}
where $i$ is an index over image pixels in both space and energy, $m_i$ indicates the number of counts predicted by the model at pixel $i$, $n_i$ is the observed number of counts at pixel $i$, and $N_{\rm exp}$ is the total number of observed counts. We rank the spatial models based on their associated likelihood statistic.

To quantify the robustness of our ranking method, we generate 10 bootstrapped realizations of our data \citetext{e.g.\ \citealp{2007nr..book.....P}}. Given the $N$ photons in the {\it Fermi} data set, our method is to select random samples of size $N$ photons allowing for replacement, i.e., some individual photons from the original data set are used more than once in a given realization, others not at all. We have released our code to generate bootstrapped samples of the \textit{Fermi} data at \url{http://github.com/garyForeman/FermiBootstrap}.

We find a wide spread in LH of the null hypothesis (i.e. the background model) among our bootstrap samples, and therefore we are not able to comment on the significance of the the likelihood statistic as a means of ranking LMC spatial models. What we can report is the ranking of each model as used for the true data set and the frequency with which the same ranking is reproduced among the bootstrapped samples. In Figures~\ref{fig:had_error} and \ref{fig:lep_error}, the error bars in the upper left plot indicate the mean and 1$\sigma$ spread of the ranks across our 10 bootstrapped samples. We will explore more advanced statistical techniques in future work.

We present the results of our binned likelihood analyses in Tables~\ref{tab:Hadron}--\ref{tab:Had+Lep}. The broad statements we can make across all of our models are as follows: (1) concentrated models, i.e., models that use star formation rate maps to represent the cosmic-ray flux, better fit the gamma-ray data than do diffuse models and lead to conclusions consistent with the conclusions of \citet{2012ApJ...750..126M} and \citet{2010A&A...512A...7A}. (2) Spectral indices fit to our subset of the LMC {\it Fermi} observations consistently lie between 2.2 and 2.4 with the exception of the 160 \m\ diffuse model, which we address below. This is the case whether 30 Dor is included or not.

\noindent
\begin{deluxetable*}{cllclcc}
\tabletypesize{\scriptsize}
\tablecolumns{8}
\tablewidth{0pc}
\tablecaption{Gas Models \label{tab:Hadron}}
\tablehead{\colhead{} & \colhead{LMC} & \colhead{$\ln(\mathrm{LH}) +$} & \colhead{LMC Flux} & \colhead{Spectral} & \colhead{$a_1$} & \colhead{$a_2$} \\
\colhead{} & \colhead{Model} & \colhead{$1.2 \times 10^5$} & \colhead{200 MeV - 20 GeV} & \colhead{Index} & \colhead{} & \colhead{} \\
\colhead{} & \colhead{} & \colhead{} & \colhead{($10^{-7}$ ph cm$^{-2}$ s$^{-1}$)} & \colhead{} & \colhead{} & \colhead{}}
\startdata
\multirow{7}{*}{With} & H$_2$ & -8250.4 & $1.58 \pm 0.03$ & $2.32 \pm 0.02$ & ... & ... \\
\multirow{7}{*}{30 Dor} & HI & -8176.8 & $1.96 \pm 0.03$ & $2.27 \pm 0.02$ & ... & ... \\
& HI+H$_2$ & -8107.2 & $1.96 \pm 0.3$ & $2.27 \pm 0.02$ & ... & ... \\
& H$_2$ ($a_1$ $\psi_{6.3}$ + $a_2$ $\psi_{12.5}$) & -7745.8 & $1.48 \pm 0.02$ & $2.34 \pm 0.02$ & 0.663 & 0.337 \\
& $\sqrt{\mathrm{H}\alpha}$ & -7569.8 & $1.91 \pm 0.03$ & $2.22 \pm 0.01$ & ... & ... \\
& H$\alpha$ & -7413.7 & $1.53 \pm 0.02$ & $2.29 \pm 0.02$ & ... & ... \\
& $\sqrt{\mathrm{H}\alpha}$ ($a_1$ $\psi_{6.3}$ + $a_2$ $\psi_{12.5}$) & -7388.7 & $1.49 \pm 0.02$ & $2.29 \pm 0.02$ & 0.204 & 0.796 \\
& HI ($a_1$ $\psi_{6.3}$ + $a_2$ $\psi_{12.5}$) & -7388.6 & $1.57 \pm 0.02$ & $2.28 \pm 0.02$ & 0.193 & 0.807 \\
& HI+H$_2$ ($a_1$ $\psi_{6.3}$ + $a_2$ $\psi_{12.5}$) & -7378.0 & $1.57 \pm 0.02$ & $2.28 \pm 0.02$ & 0.207 & 0.793 \\
\hline
\multirow{7}{*}{Without} & H$_2$ & -7750.8 & $1.11 \pm 0.03$ & $2.34 \pm 0.02$ & ... & ... \\
\multirow{7}{*}{30 Dor} & HI & -7700.6 & $1.40 \pm 0.03$ & $2.28 \pm 0.02$ & ... & ... \\
& HI+H$_2$ & -7673.9 & $1.41 \pm 0.03$ & $2.27 \pm 0.02$ & ... & ... \\
& H$_2$ ($a_1$ $\psi_{6.3}$ + $a_2$ $\psi_{12.5}$) & -7631.3 & $1.02 \pm 0.03$ & $2.34 \pm 0.02$ & 0.569 & 0.431 \\
& $\sqrt{\mathrm{H}\alpha}$ & -7383.7 & $1.40 \pm 0.03$ & $2.22 \pm 0.02$ & ... & ... \\
& H$\alpha$ & -7336.9 & $1.25 \pm 0.03$ & $2.26 \pm 0.02$ & ... & ... \\
& $\sqrt{\mathrm{H}\alpha}$ ($a_1$ $\psi_{6.3}$ + $a_2$ $\psi_{12.5}$) & -7532.5 & $1.02 \pm 0.03$ & $2.34 \pm 0.02$ & 1.000 & 0.000 \\
& HI ($a_1$ $\psi_{6.3}$ + $a_2$ $\psi_{12.5}$) & -7409.3 & $1.14 \pm 0.03$ & $2.26 \pm 0.04$ & 0.209 & 0.791 \\
& HI+H$_2$ ($a_1$ $\psi_{6.3}$ + $a_2$ $\psi_{12.5}$) & -7395.6 & $1.14 \pm 0.03$ & $2.26 \pm 0.02$ & 0.205 & 0.795 \\
\vspace{-10pt}
\enddata
\tablecomments{Models ordered by their with 30 Dor ranking based on $\ln(\mathrm{LH})$. Here, $a_1$ and $a_2$ represent the fraction of flux contributed by the 6.3 and 12.5 Myr star formation rate maps respectively.}
\end{deluxetable*}

\begin{deluxetable*}{clcclcc}
\tabletypesize{\scriptsize}
\tablecolumns{8}
\tablewidth{0pc}
\tablecaption{Radiation Models \label{tab:Lepton}}
\tablehead{\colhead{} & \colhead{LMC} & \colhead{$\ln(\mathrm{LH}) +$} & \colhead{LMC Flux} & \colhead{Spectral} & \colhead{$a_1$} & \colhead{$a_2$} \\
\colhead{} & \colhead{Model} & \colhead{$1.2 \times 10^5$} & \colhead{200 MeV - 20 GeV} & \colhead{Index} & \colhead{} & \colhead{} \\
\colhead{} & \colhead{} & \colhead{} & \colhead{($10^{-7}$ ph cm$^{-2}$ s$^{-1}$)} & \colhead{} & \colhead{} & \colhead{}}
\startdata
\multirow{6}{*}{With} & 1.4 GHz $\times$ 160 \m & -8651.4 & $0.92 \pm 0.02$ & $2.52 \pm 0.03$ & ... & ... \\
\multirow{6}{*}{30 Dor} & 1.4 GHz $\times$ 1.24 \m & -7862.4 & $1.34 \pm 0.02$ & $2.29 \pm 0.02$ & ... & ... \\
& 1.24 \m ($a_1$ $\psi_{6.3}$ + $a_2$ $\psi_{12.5}$) & -7569.7 & $1.46 \pm 0.02$ & $2.31 \pm 0.02 $ & 0.977 & 0.033 \\
& CMB ($a_1$ $\psi_{6.3}$ + $a_2$ $\psi_{12.5}$) & -7549.1 & $1.63 \pm 0.03$ & $2.26 \pm 0.02$ & 0.839 & 0.161 \\
& 1.4 GHz $\times$ CMB & -7463.1 & $1.60 \pm 0.03$ & $2.26 \pm 0.02$ & ... & ... \\
& 160 \m ($a_1$ $\psi_{6.3}$ + $a_2$ $\psi_{12.5}$) & -7408.0 & $1.51 \pm 0.02$ & $2.28 \pm 0.02$ & 0.174 & 0.826 \\
\hline
\multirow{6}{*}{Without} & 1.4 GHz $\times$ 160 \m & -7595.2 & $1.23 \pm 0.03$ & $2.34 \pm 0.02$ & ... & ... \\
\multirow{6}{*}{30 Dor} & 1.4 GHz $\times$ 1.24 \m & -7895.6 & $0.94 \pm 0.03$ & $2.29 \pm 0.03$ & ... & ... \\
& 1.24 \m ($a_1$ $\psi_{6.3}$ + $a_2$ $\psi_{12.5}$) & -7513.1 & $1.05 \pm 0.03$ & $2.31 \pm 0.02$ & 1.00 & 0.00 \\
& CMB ($a_1$ $\psi_{6.3}$ + $a_2$ $\psi_{12.5}$) & -7396.0 & $1.11 \pm 0.03$ & $2.25 \pm 0.02$ & 0.391 & 0.609 \\
& 1.4 GHz $\times$ CMB & -7460.6 & $1.34 \pm 0.03$ & $2.27 \pm 0.02$ & ... & ... \\
& 160 \m ($a_1$ $\psi_{6.3}$ + $a_2$ $\psi_{12.5}$) & -7611.1 & $0.98 \pm 0.03$ & $2.36 \pm 0.03$ & 0.588 & 0.412 \\
\vspace{-10pt}
\enddata
\tablecomments{Models ordered by their with 30 Dor ranking based on $\ln(\mathrm{LH})$. Here, $a_1$ and $a_2$ represent the fraction of flux contributed by the 6.3 and 12.5 Myr star formation rate maps respectively.}
\label{tab:Lepton}
\end{deluxetable*}

\begin{deluxetable*}{clccccc}
\tabletypesize{\scriptsize}
\tablecolumns{7}
\tablewidth{0pc}
\tablecaption{Gas + Radiation Model \label{tab:Had+Lep}}
\tablehead{\colhead{} & \colhead{LMC} & \colhead{$\ln(\mathrm{LH}) +$} & \colhead{LMC Flux} & \colhead{Spectral} & \colhead{$a_1$} & \colhead{$a_2$} \\
\colhead{} & \colhead{Model} & \colhead{$1.2 \times 10^5$} & \colhead{200 MeV - 20 GeV} & \colhead{Index} & \colhead{} & \colhead{} \\
\colhead{} & \colhead{} & \colhead{} & \colhead{($10^{-7}$ ph cm$^{-2}$ s$^{-1}$)} & \colhead{} & \colhead{} & \colhead{}}
\startdata
\multirow{1}{*}{With} & $a_1 \sqrt{\rm H\alpha}$ + & -7307.4 & $1.72 \pm 0.08$ & $2.22 \pm 0.05$ & 0.449 & 0.551 \\
\multirow{1}{*}{30 Dor} & $a_2$ 160 \m ($C \psi_{6.3}$ + ($1 - C$) $\psi_{12.5}$) & ... & ... & $2.26 \pm 0.03$ & ... & ... \\
\hline
\multirow{1}{*}{Without} & $a_1 \sqrt{\rm H\alpha}$ + & -7321.9 & $1.35 \pm 0.06$ & $2.21 \pm 0.03$ & 0.761 & 0.239 \\
\multirow{1}{*}{30 Dor} & $a_2$ 160 \m ($C \psi_{6.3}$ + ($1 - C$) $\psi_{12.5}$) & ... & ... & $2.28 \pm 0.07$ & ... & ... \\
\vspace{-10pt}
\enddata
\tablecomments{Likelihood statistic reported here not directly comparable to Tables~\ref{tab:Hadron} and \ref{tab:Lepton} because more free parameters are used in these models. Here, $a_1$ and $a_2$ represent the fraction of flux contributed by the gas and radiation maps respectively. $C$ and $(1 - C)$ correspond to $a_1$ and $a_2$ from Table~\ref{tab:Lepton}.}
\end{deluxetable*}

The total LMC flux $>$200 MeV estimated from our models range from $0.92 \times 10^{-7}$ to $1.96 \times 10^{-7}$ ph cm$^{-2}$ s$^{-1}$ for the cases in which we include the 30 Doradus data, and fluxes range from $0.94 \times 10^{-7}$ to $1.40 \times 10^{-7}$ ph cm$^{-2}$ s$^{-1}$ when we exclude 30 Doradus. As expected, for each given model, estimated fluxes are higher when we include the 30 Doradus data, with the one anomaly being the diffuse 160~\m\ radiation model. Spectral indices range from 2.22 to 2.52.

Model-to-model changes in the LMC flux values are mirrored by compensating changes in the estimated Milky Way flux. The difference between the LMC and the Milky Way in the absolute variation in flux is explained by the fact that the Milky Way's flux is across the entire sky, whereas our region of interest is $\sim$1\% of the sky. The flux differences between models with 30 Dor and those without are well explained by  the 30 Dor point source flux in conjunction with the Milky Way flux.

{\begin{figure*}[ht]
\centering
\includegraphics[width=6.6in]{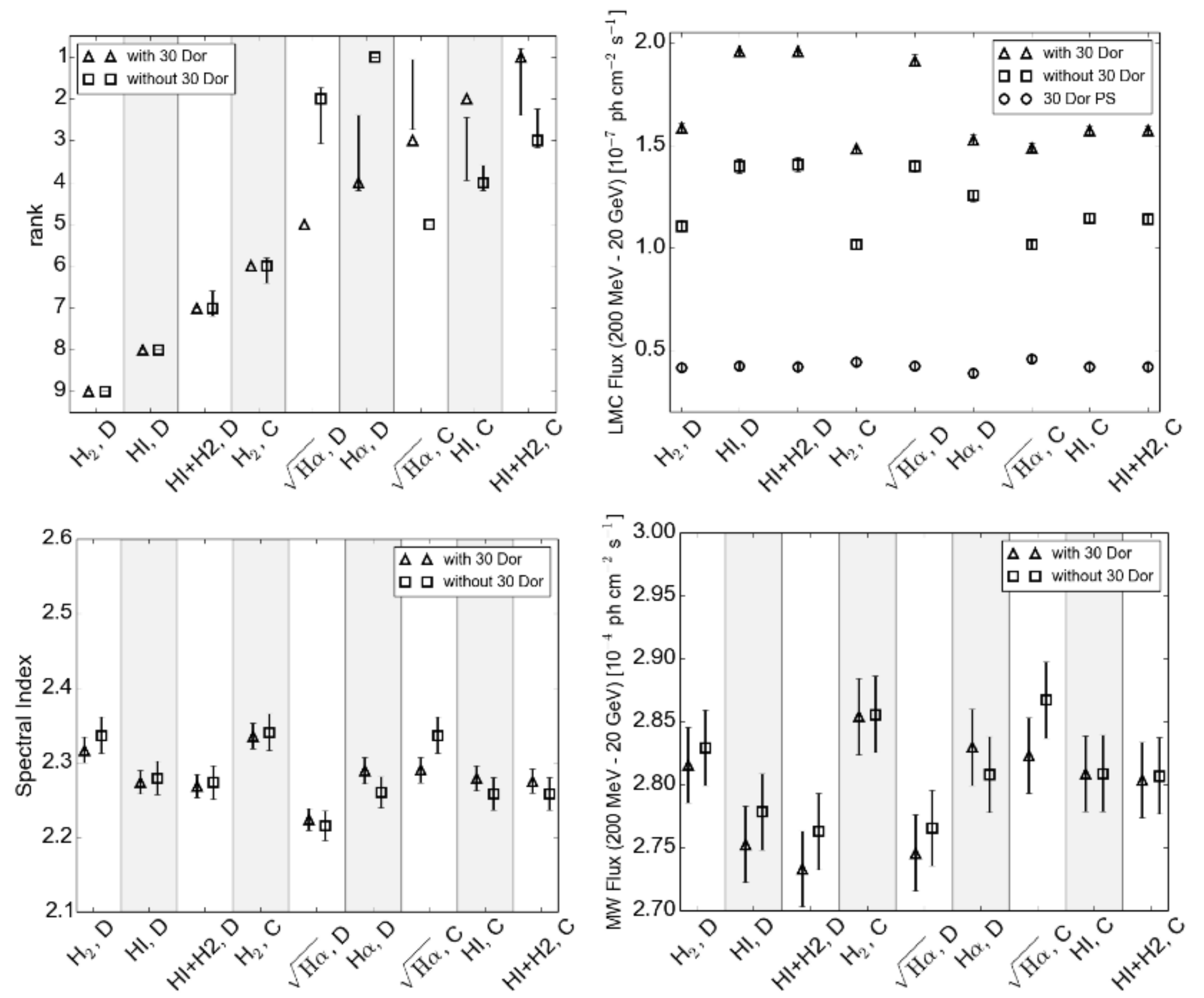}
\figcaption{Bootstrapped results for the gas models. Triangles (full LMC) and squares (30 Dor removed) indicate the values found from the {\it Fermi} data set (data listed in Table~\ref{tab:Hadron}). The error bars indicate a $\pm 1\sigma$ spread on the mean quantities found from ten bootstrapped samples. We have slightly offset the full LMC data points from the 30 Dor removed data points to avoid error bar confusion. Gray shaded regions are meant to help distinguish between quantities from separate models. {\it Upper left}: rank as determined by the likelihood statistic. {\it Upper right}: LMC flux. {\it Lower left}: spectral index. {\it Lower right}: Milky Way foreground flux. \label{fig:had_error}}
\end{figure*}}

{\begin{figure*}[ht]
\centering
\includegraphics[width=6.6in]{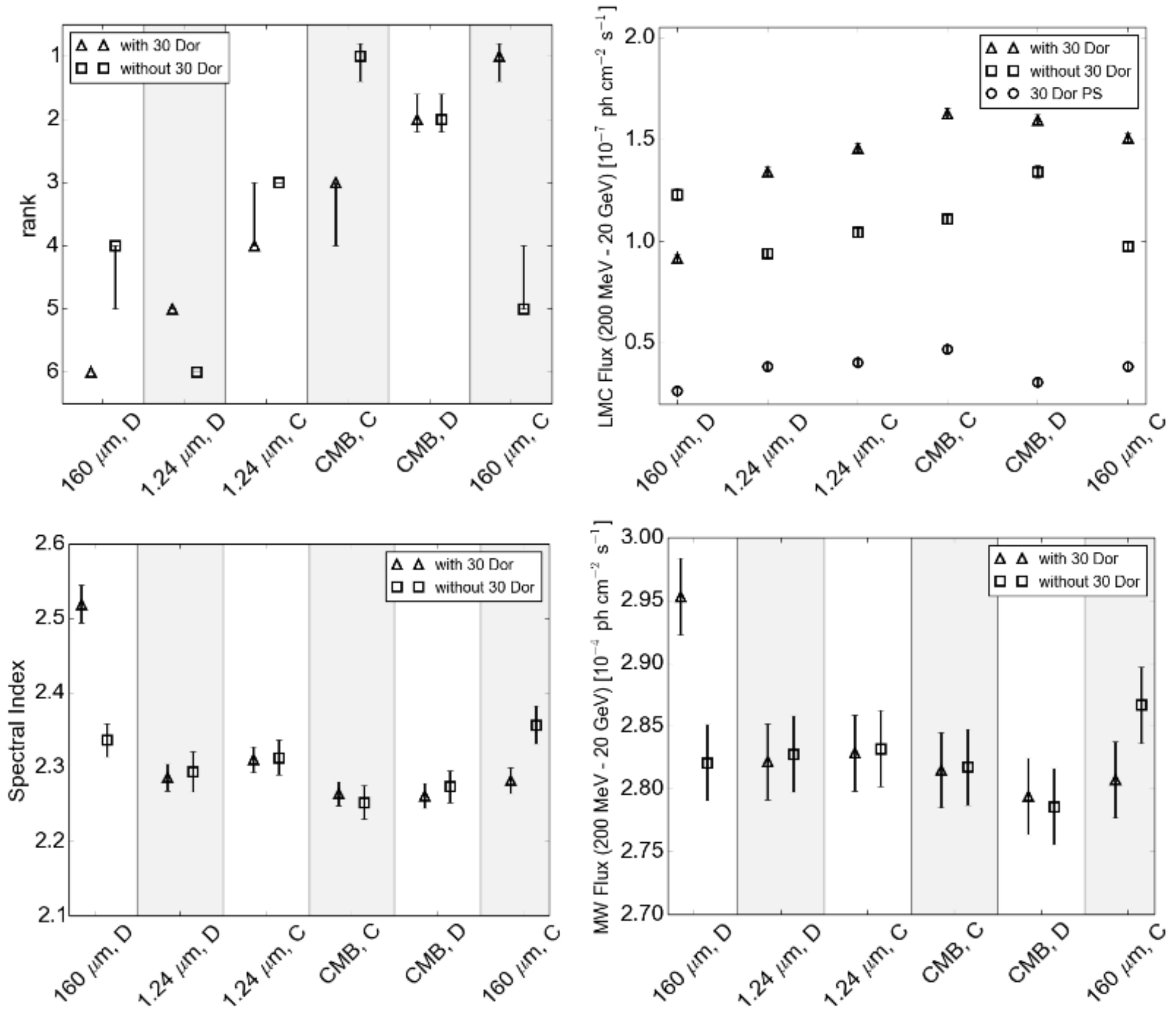}
\figcaption{Same as Figure~\ref{fig:had_error}, but for radiation models. Note: the best fitting gas model, ``HI+H2, C,'' consistently outranks the ``160 \m, C'' model. \label{fig:lep_error}}
\end{figure*}}

\subsection{Gas Models}
As did \citet{2010A&A...512A...7A}, we find the H$\alpha$ model consistently outranks the HI and H$_2$ models both when including and excluding the 30 Dor region. However, we find that that our optically thin HI model better fits the gamma-ray data than H$_2$ whereas the {\it Fermi} study found the inverse. We are unfortunately precluded from making a more quantitative comparison because \citet{2010A&A...512A...7A} do not report the normalization and spectral index parameters estimated from their maximum likelihood fits of these spatial templates. Our stricter requirements on photon quality and the extra point sources we have used in our background model exacerbate this difficulty.

The upper-left plot of Figure~\ref{fig:had_error} reveals that when 30 Dor is included, concentrated models are consistently ranked higher than their diffuse counterparts. We remind the reader that diffuse gas models assume a spatially uniform cosmic-ray flux distribution, whereas concentrated models approximate the cosmic-ray flux using the star formation rate maps of Figure~\ref{fig:HZSF}. The four best-fitting models are indistinguishable based on our bootstrapped ranking errors, but, again, we do not consider the diffuse \ha\ model to be a good estimate of the ionized hydrogen column density. This model has been included only for comparison with \citet{2010A&A...512A...7A}.

When we exclude the 30 Dor region, we find the diffuse H$\alpha$ and $\sqrt{\rm H\alpha}$ maps to consistently outrank all concentrated models. We remind the reader that we do not consider the H$\alpha$ model to be physical because H$\alpha$ intensity is not proportional to the ionized hydrogen column density. However, the high ranking of the $\sqrt{\rm H\alpha}$ does stand out, especially when comparing its $\ln(\mathrm{LH})$ value with the other models (see Table~\ref{tab:Hadron}).

The LMC flux varies by $\sim$25\% depending on which spatial model is used to to estimate the flux. The high fluxes from the diffuse HI and diffuse HI+H2 models can be understood by the fact that the the HI emission extends farther from the LMC center than do emissions from other components. Spectral indices vary at about 5\% depending on the spatial model, and the estimated Milky Way fluxes vary at about 4\%.

\subsection{Radiation Models}
For the radiation models plotted in Figure~\ref{fig:lep_error}, the statement that concentrated models better explain the LMC gamma-ray data is less strong. This is a reflection of the different diffuse model we use to represent the cosmic-ray electron flux. Rather than assuming a spatially uniform distribution, we use the synchrotron radiation map presented in Figure~\ref{fig:nthRadio}. The synchrotron map traces low-energy cosmic-ray electrons, some of which have lost energy through inverse Compton scattering. These older electrons have had more time to diffuse throughout the LMC than their younger, high-energy counterparts, so we use them for our diffuse model.

The diffuse 160 \m\ model exhibits strange behavior in that it becomes brighter when the 30 Dor photons are removed. This finding is explained by the fact that this model is highly peaked in the 30 Dor region, which in turn dominates the fit. The rest of the diffuse emission is under-explained by this model and is ultimately incorporated into the fit of the diffuse Milky Way model (see lower right plot of Figure~\ref{fig:lep_error}). When the peaked 30 Dor region is removed, the rest of the LMC photons are better explained by the model resulting in a higher LMC flux estimate.

The spread in estimated model parameters from the radiation models is heavily influenced by the diffuse 160 \m\ model, which we consider anomalous. If we include the 160 \m\ model, the spread in LMC flux is about 44\%; the spread in spectral index is $\sim$10\%; and the spread in Milky Way flux is $\sim$5\%. These spreads reduce to respectively 18\%, 5\%, and 3\% if we discount the diffuse 160 \m\ model.

Finally, we wish to note that the concentrated HI+H$_2$ model consistently outranks the best fitting radiation model.

\subsection{Gas + Radiation Models}
As mentioned in \S\ref{sec:had+lep}, we have chosen several combinations of best fitting gas and radiation models to be fit simultaneously. Among these models, we find the inverse Compton contribution to the flux to be generally between 30\% and 50\%. We choose one fiducial model to present here, which is the diffuse $\sqrt{\rm H\alpha}$ model combined with the concentrated 160 \m. This model consistently outranked other combinations. Results of the model fits are listed in Table~\ref{tab:Had+Lep}.

This model estimates the inverse Compton flux contribution at 55\% when the 30 Dor region is included and at 24\% when 30 Dor is excluded. The spectral indices fit to the gas and radiation components are consistent with each other for both data sets. Note the improvement in likelihood of the combined model compared with the diffuse $\sqrt{\rm H\alpha}$ model. This likelihood improvement accompanies a lower estimated LMC flux for the gas + radiation model.

In summary, we find that the best fit gas model consistently outranks the best fit radiation model. However, when we combine gas and radiation models and fit simultaneously, we find significant flux is contributed by the radiation component, which suggests the leptonic contribution to the LMC gamma-ray signal should not be ignored (further discussion in \S~\ref{sec:LeptonEmission}). We also find concentrated models generally outperform their diffuse counterparts, favoring short diffusion scale lengths (see \S~\ref{sec:Diffusion}).

\section{SPECTRAL ANALYSIS} \label{sec:Spectral}
In this section, we consider the spatially integrated LMC gamma-ray spectrum from the data presented in Figure~\ref{fig:binned_counts}. We split our data into six logarithmically spaced energy bins from 200 MeV to 20 GeV and repeat the binned likelihood analysis on each bin independently. To estimate the gamma-ray flux from each bin, we use the diffuse $\sqrt{\rm H\alpha}$ +  concentrated 160~\m\ model. In this analysis, we model the LMC gamma-ray emission with twelve free parameters: two normalizations per energy bin, where the spectral index within each energy bin has been fixed to zero. As for the LMC models, all point sources have been fixed to have a spectral index of zero.

Performing our analysis on individual energy bins in this way allows us to remove the assumption of a single power law used in our spatial analysis. This methodology also allows us to investigate the pion decay, bremsstrahlung, and inverse Compton flux contributions by fitting theoretical curves to the LMC gamma-ray spectrum. \citet{2004A&A...413...17P} give a fitting function for the gamma-ray spectrum predicted by pion decay:
\begin{equation}
\label{eq:pionBump}
\begin{split}
\frac{d N_{\gamma}}{dA\ dt\ dE\ d\Omega} =& \frac{1}{4 \pi}\sigma_{\rm pp} c N_{\rm H} 2^{2 - \alpha} \frac{\tilde{n}_{\rm CRp}}{\rm GeV} \frac{4}{3 \alpha} \left(\frac{m_{\pi^0} c^2}{\rm GeV}\right)^{-\alpha} \times \\
& \left[\left(\frac{2 E_{\gamma}}{m_{\pi^0} c^2}\right)^{\delta} + \left(\frac{2 E_{\gamma}}{m_{\pi^0} c^2}\right)^{-\delta}\right]^{\alpha / \delta}.
\end{split}
\end{equation}
The first free parameter of this function is a normalization factor that is proportional to $N_{\rm H} \tilde{n}_{\rm CRp}$, the hydrogen column density times a proxy for the number density of cosmic-ray protons above the energy threshold for neutral pion production ($E_{\rm thr} = 1.2$ GeV). The second free parameter is the high-energy power law index, $\alpha$. The proton-proton collision cross section is given as $\sigma_{\rm pp} = 32 (0.96 + e^{4.4 - 2.4 \alpha})$ mbarn, and the exponent $\delta = 0.14 \alpha^{-1.6} + 0.44$.

Similarly, \citet{2013ApJ...773..104C} give an inverse Compton fitting function for normal, i.e., non starbursting, galaxies:
\begin{equation}
\label{eq:inverseCompton}
\begin{split}
&\log_{10}\left(E^2 \frac{dN_{\gamma}}{dt\ dE}\right) = \log_{10}\left(\frac{\psi}{\psi_{\rm MW}}\right) - \\
& \indent \left[-40.26 + X(0.0949 + X(0.0503 + 0.0251X))\right],
\end{split}
\end{equation}
where $X = \log_{10}(E / {\rm GeV})$, and $\psi$ is the star formation rate. The functional form of this inverse Compton spectrum is based on a one-zone model of the Milky Way that assumes electron calorimetry, which leads to linear scaling with star formation rate. Equation~(\ref{eq:inverseCompton}) is normalized to the total Milky Way inverse Compton luminosity found in the GALPROP plain diffusion model \citep{2010ApJ...722L..58S}. Because of this normalization, Equation~\ref{eq:inverseCompton} implicitly assumes the same cosmic-ray acceleration efficiency. Other cosmic ray models (e.g., diffusive reacceleration, different confinement volumes) can lead to different efficiencies. Equation~(\ref{eq:inverseCompton}) also assumes that the galaxy with star formation rate $\psi$ has the same ratio of electron energy-loss rates (i.e.\ $b_{\rm IC} / b_{\rm sync}$ and $b_{\rm IC} / b_{\rm brem}$) as the Milky Way. This is not true of the LMC, which has a much higher neutral hydrogen density and a lower ISRF energy density. Therefore, we scale Equation~(\ref{eq:inverseCompton}) by an additive factor of 
\begin{equation}\label{eq:ICscaleFactor}
\log_{10}\left(\frac{U_{\rm LMC, ISRF}}{U_{\rm MW, ISRF}} \frac{b_{\rm tot, MW}(\rm 1\ GeV)}{b_{\rm tot, LMC}(\rm 1\ GeV)}\right) \sim -1,
\end{equation}
where $U_{\rm ISRF}$ is the energy density of the ISRF, and $b_{\rm tot}$ is the sum of four electron energy loss rates: $b_{\rm sync}$, $b_{\rm IC}$, $b_{\rm brem}$, and $b_{\rm ion}$, where $b_{\rm ion}$ is the loss rate from ionization \citep{1964ocr..book.....G}. For further details about the quantities we have used for the LMC, please refer to Appendix~\ref{app:bremSpec}. We adopt $\psi_{\rm MW}$, $U_{\rm MW, ISRF}$, and $b_{\rm tot, MW}(\rm 1\ GeV)$ from the fiducial \citet{2013ApJ...773..104C} model. The scaling of Equation~(\ref{eq:ICscaleFactor}) is motivated by the unnumbered relation that appears between Eqs.~(23) and (24) of \citet{2013ApJ...773..104C}. 

Because of the calorimetry of inverse Compton emission, Equation~(\ref{eq:inverseCompton}) rescaled by (\ref{eq:ICscaleFactor}) gives a zero parameter prediction for the LMC inverse Compton luminosity, if we assume a universal cosmic-ray acceleration efficiency. The detailed shape of the inverse Compton spectrum depends on the details of the LMC ISRF, but as shown in \citet{2013ApJ...773..104C}, in the calorimetric limit, the dependence is mild for a plausible range of radiation field components (UV, optical, IR, and CMB). Alternatively, if one adopts the observed LMC star formation rate, then the inverse Compton luminosity normalization measures the ratio $\epsilon_{e,\rm LMC}/\epsilon_{e,\rm MW}$ of cosmic-ray electron efficiency in the LMC to that in the Milky Way.

We have formulated our own bremsstrahlung fitting function based on the same one-zone model discussed in the previous paragraph, which is a modified version of the \citet{2013ApJ...773..104C} model. The detailed derivation of this fitting function can be found in Appendix~\ref{app:bremSpec}. The equation we fit takes the final form
\begin{equation}\label{eq:bremsstrahlung}
\begin{split}
& \frac{dN_{\gamma}}{dt\ dE} = 5.93 \times 10^{29}\ {\rm s^{-1}}\ \frac{\psi_{\rm LMC}}{\psi_{\rm MW}}\frac{4 \alpha}{\pi} \sigma_T \ln(183) n_{\rm H} c E_{\gamma}^{-1} \\
& \indent \times \left\{
\begin{array}{ll}
\int_{E_{\gamma}} \frac{dE_e}{b(E_e)}[1.25 (E_e^{-0.8} - E_b^{-0.8}) & \\
\indent + 0.8 (E_b^{-1.25} - E_{\rm max}^{-1.25})] & E_{\gamma} < E_b \\
\int_{E_{\gamma}} \frac{dE_e}{b(E_e)}[0.8 (E_e^{-1.25} - E_{\rm max}^{-1.25})] & E_b < E_{\gamma} < E_{\rm max}
\end{array}
\right. \\
\end{split}.
\end{equation}
This spectrum is based on a broken power-law spectrum of the cosmic-ray electron energies $E_e$ with break energy $E_b = 4$ GeV and a hard cutoff at $E_{\rm max} = 2$ TeV. Here $\alpha$ is the fine structure constant, $\sigma_T$ is the Thomson scattering cross section, $n_{\rm H}$ is the total hydrogen density, and $b(E_e)$ is the sum of electron energy-loss rates from bremsstrahlung, inverse Compton scattering, synchrotron, and ionization.

The reason bremsstrahlung and inverse Compton spectra depend on the star formation ratio is that the cosmic-ray electron injection rate is proportional to the supernova rate, which is proportional to the star formation rate. Both Eqs.~(\ref{eq:inverseCompton}) and Eqs.~(\ref{eq:bremsstrahlung}) have been normalized to the Milky Way spectra calculated from GALPROP \citep{2010ApJ...722L..58S}. Therefore, to apply these spectra to other galaxies with different cosmic-ray densities, one scales by the star formation rate.

Ideally, we could constrain the free parameters of Eqs.~(\ref{eq:pionBump}), (\ref{eq:inverseCompton}), and (\ref{eq:bremsstrahlung}) using the maximum likelihood estimator implemented in the {\it Fermi} Science Tools software. Unfortunately, the normalization is the only free parameter allowed for user-defined spectral functions, so we would not be able to estimate the high-energy spectral index of Equation~(\ref{eq:pionBump}). We proceed by using least squares to fit these spectral profiles to the binned flux data. The contribution to the fit from each energy bin is weighted by the flux error.

Figure~\ref{fig:SpecAnalysis} shows the results of the fits described above. Each plot shows a different subset of the LMC data. The solid lines indicate the results of the simultaneous fits of Eqs.~(\ref{eq:pionBump}), (\ref{eq:inverseCompton}), and (\ref{eq:bremsstrahlung}). The dashed, dotted, and dotted-dashed lines show the contributions from the pion bump, bremsstrahlung, and inverse Compton models respectively. We present the fitted parameters in Table~\ref{tab:specFits}.  

{\begin{figure}[!t]
\centering
\includegraphics[width=3in]{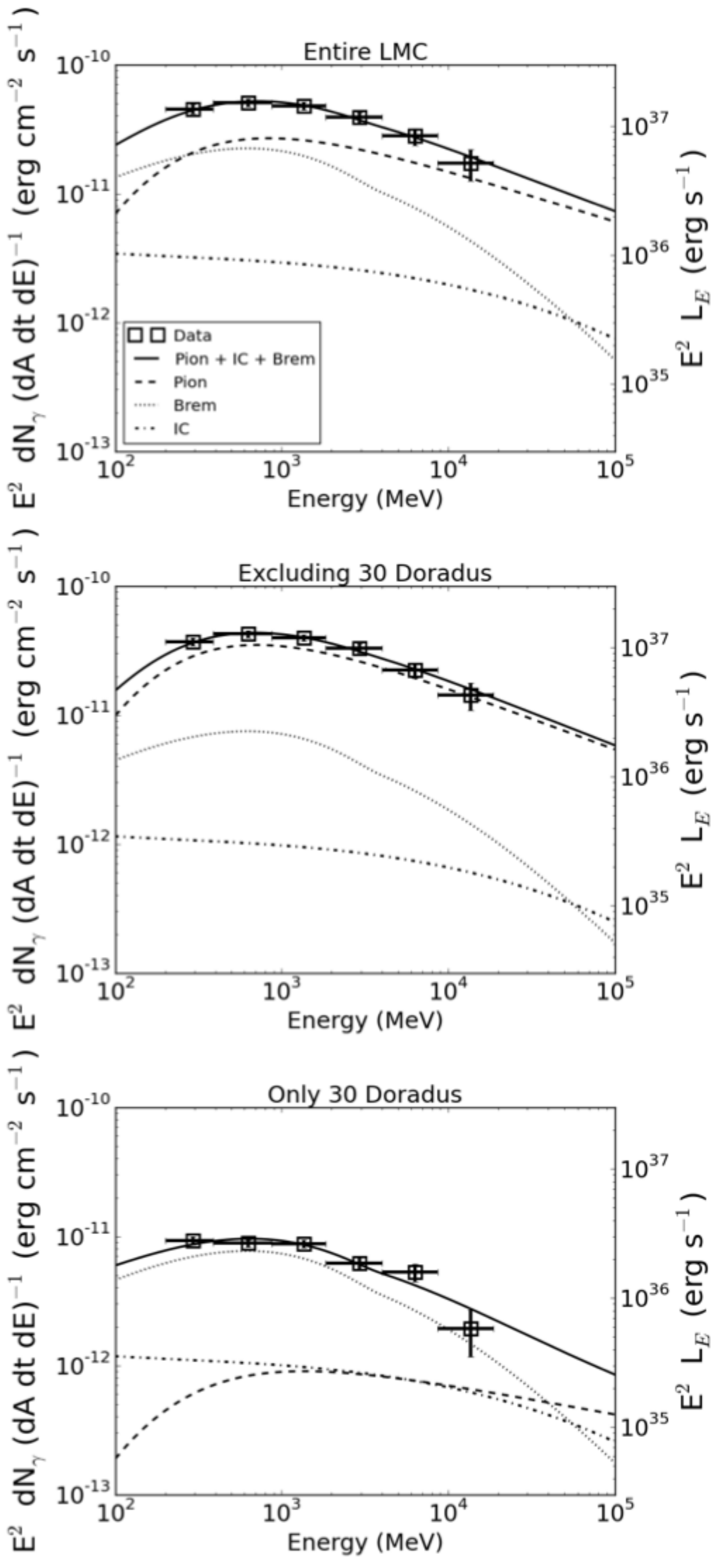}
\figcaption{Spectral fits to the binned flux estimates. The fit is performed using Equations~(\ref{eq:pionBump}) + (\ref{eq:inverseCompton}) and three free parameters are fit simultaneously: $N_{\rm H} \tilde{n}_{\rm CRp}$, $\alpha$, and $\psi_{\rm LMC} / \psi_{\rm MW}$. Solid lines show the result of the fits, and the dashed, dotted, and dot-dashed lines show the contributions from the pion bump, bremsstrahlung, and inverse Compton models respectively. {\it Top}: model for the full LMC. {\it Center}: model for the LMC data excluding 30 Dor. {\it Bottom}: point source model for 30 Dor. \label{fig:SpecAnalysis}}
\end{figure}}

\begin{deluxetable*}{lcccccccc}
\tabletypesize{\scriptsize}
\tablecolumns{7}
\tablewidth{0pt}
\tablecaption{Spectral Fits \label{tab:specFits}}
\tablehead{\colhead{Data Set} & \colhead{$N_{\rm H} \tilde{n}_{\rm CRp}$} & \colhead{max$(N_{\rm H} \tilde{n}_{\rm CRp})$} & \colhead{$\alpha$} & \colhead{$\psi_{\rm LMC} / \psi_{\rm MW}$} & \colhead{max$(\psi_{\rm LMC} / \psi_{\rm MW})$} & \colhead{Pion Flux} & \colhead{Brem Flux} & \colhead{IC Flux}}
\startdata
Full LMC & 1.3 & 3.4 & $2.4 \pm 0.3$ & 0.3 & 0.7 & 6.9 & 6.1 & 0.9 \\
No 30 Dor & 2.1 & 2.8 & $2.5 \pm 0.2$ & 0.1 & 0.6 & 9.1 & 2.0 & 0.3 \\
30 Dor Point Source & 0.03 & 0.75 & $2.2 \pm 1.9$ & 0.1 & 0.1 & 0.2 & 2.0 & 0.3 \\
\enddata
\vspace{-10pt}
\tablecomments{$N_{\rm H} \tilde{n}_{\rm CRp}$ given in units of 10$^{11}$ cm$^{-5}$. Fluxes given in units of 10$^{-8}$ ph cm$^{-2}$ s$^{-1}$. max($N_{\rm H} \tilde{n}_{\rm CRp}$) computed by setting $\psi_{\rm LMC} / \psi_{\rm MW} = 0$. max($\psi_{\rm LMC} / \psi_{\rm MW}$) computed by setting $N_{\rm H} \tilde{n}_{\rm CRp} = 0$.}
\end{deluxetable*}

The LMC gamma-ray flux from 200 MeV to 20 Gev is $(1.37 \pm 0.02) \times 10^{-7}$ ph cm$^{-2}$ s$^{-1}$, which we estimate by integrating the fitted spectral function indicated by the solid curve in the top panel of Figure~\ref{fig:SpecAnalysis}. The error bounds are calculated by performing the least squares fit on an ensemble of bootstrapped samples.

We can estimate the flux contributions from each of the gamma-ray production channels. For the full LMC data set, our spectral analysis suggests that 6\% of the LMC gamma-ray flux is caused by inverse Compton scattering, and 44\% is from bremsstrahlung. For regions outside of 30 Dor, our spectral fit estimates that 3\% of the gamma-ray flux is from inverse Compton and 18\% is from bremsstrahlung. For the 30 Dor point source model, the spectral analysis results in an inverse Compton flux contribution of 12\% and a bremsstrahlung contribution of 80\%.

As indicated in Table~\ref{tab:specFits}, the $N_{\rm H} \tilde{n}_{\rm CRp}$ and the $\psi_{\rm LMC} / \psi_{\rm MW}$ parameters are not particularly well constrained. The reason the errors on the spectral flux estimate are so small in spite large variation in the fitting parameters is that a Pearson correlation coefficient of $r = -0.998$ indicates $N_{\rm H} \tilde{n}_{\rm CRp}$ and $\psi_{\rm LMC} / \psi_{\rm MW}$ are nearly perfectly anticorrelated. This is caused by the fact that the sum of the bremsstrahlung and inverse Compton spectra as parameterized by $\psi_{\rm LMC} / \psi_{\rm MW}$ has such a similar shape to that of the pion spectrum parameterized by $N_{\rm H} \tilde{n}_{\rm CRp}$.

The high-energy power-law index is better constrained than the two normalization parameters and consistent across all three data sets. An index $\alpha = 2.4$ is consistent with the result of all three fits. This finding has a strong implication for the underlying cosmic-ray proton spectrum itself as we discuss below. As mentioned above, the spectral indices found from our spatial fits, which have a typical value of about 2.3, slightly underestimate the high energy power law index $\alpha$ because the the spatial models assumed a power law over the entire energy range.

\section{DISCUSSION}\label{sec:Discussion}
\subsection{Cosmic-ray Density and Magnetic Field Strength}
At high energies, the $\pi^0$ spectrum (and by extension the gamma-ray spectrum) is a power law with index equal to that of the cosmic-ray protons that create them \citetext{e.g., \citealp{1971NASSP.249.....S}; \citealp{1986A&A...157..223D}; \citealp{1991crpp.book.....G}}. This means the $\alpha$ parameter in Equation~(\ref{eq:pionBump}) is itself the cosmic-ray proton spectral index. By fitting Equation~(\ref{eq:pionBump}) to the energy-binned gamma-ray data of the entire LMC, we find $\tilde{n}_{\rm CRp} N_{\rm H} = 1.3 \times 10^{11}$ cm$^{-5}$, and the power-law index of the cosmic-ray proton population is $\alpha = 2.4$. \citet{2010A&A...512A...7A} take $\int N_{\rm H} d \Omega = 3.6 \times 10^{19}$ H-atom cm$^{-2}$ sr as an average of measurements made by \citet{2003ApJS..148..473K} and \citet{bernardetal08}. This result corresponds to $N_{\rm H} = 1.4 \times 10^{21}$ H-atom cm$^{-2}$, assuming an angular size of 81 deg$^2$ for the LMC (size of Figures~\ref{fig:hadron}, \ref{fig:ISRF}, and \ref{fig:HZSF}). From our fit, $\tilde{n}_{\rm CRp} = 9.3 \times 10^{-11}$ cm$^{-3}$.

From $\tilde{n}_{\rm CRp}$ and $\alpha$, \citet{2003A&A...407L..73P} give the cosmic-ray proton kinetic energy density as
\begin{equation}
\begin{split}
\label{eq:energyDensity}
\varepsilon_{\rm CRp} =& \frac{\tilde{n}_{\rm CRp} m_{\rm p} c^2}{2 (\alpha -1)} \left(\frac{m_{\rm p} c^2}{\rm GeV}\right)^{1 - \alpha} \times \\
& \left[\mathcal{B}_x\left(\frac{\alpha - 2}{2},\ \frac{3 - \alpha}{2}\right) + 2 \tilde{p}^{1 - \alpha} \left(\sqrt{1 + \tilde{p}^2} - 1\right)\right].
\end{split}
\end{equation}
Here, $\mathcal{B}_x(a, b)$ is the incomplete beta function, $x = (1 + \tilde{p}^2)^{-1}$, and $\tilde{p} = p_{\rm min} / (m_{\rm p} c)$. We take $p_{\rm min}$ to be the minimum cosmic-ray proton momentum necessary for pion production, which is 780 MeV c$^{-1}$ calculated from relativistic proton collision kinematics. From Equation~(\ref{eq:energyDensity}), we find the cosmic-ray proton energy density to be $3.1 \times 10^{-13}$ erg cm$^{-3}$. If we assume the energy density of cosmic rays is dominated by protons and that magnetic fields and cosmic rays are in energy equipartition, the result corresponds to a magnetic field strength of 2.8 $\mu$G.

Our magnetic field strength estimate is in good agreement with \citet{2005Sci...307.1610G}, who found the field strength to be 4.3 $\mu$G based on Faraday rotation measurements of the LMC. \citet{2012ApJ...759...25M} also calculated an equipartition-based magnetic field strength of $\sim$2 $\mu$G based on the results from \citet{2010A&A...512A...7A}. In their calculation, \citet{2012ApJ...759...25M} used a harder spectral index of $\alpha = 2.0$, which was found by \citet{2010A&A...512A...7A} by fitting a power law plus exponential cutoff to the gamma-ray spectrum. This means the power-law index they report fits the {\it low}-energy end of the spectrum, the regime of the pion bump turnover. As mentioned above, the {\it high}-energy portion of the gamma-ray spectrum is determined by the cosmic-ray proton spectrum, which is better modeled by the \citet{2004A&A...413...17P} fitting function, Equation~(\ref{eq:pionBump}).

Despite the discrepancy in details, we are able to draw the same conclusions from our magnetic field calculation as \citet{2012ApJ...759...25M}. Those authors made a second estimate of $\sim$7 $\mu$G using radio synchrotron data. The agreement between the equipartition and synchrotron estimates along with the measurement reported by \citet{2005Sci...307.1610G} led \citet{2012ApJ...759...25M} to conclude the cosmic-ray energy equipartition with magnetic fields is not violated in the LMC.

\subsection{Leptonic Gamma-ray Emission in the LMC} \label{sec:LeptonEmission}
The previous studies of the the {\it Fermi} LMC data \citep{2010A&A...512A...7A, 2012ApJ...750..126M} assumed that the gamma-ray emission is dominated by hadronic collisions and subsequent neutral pion decay. Both our spatial and spectral analyses suggest leptonic processes contribute significantly to the LMC gamma-ray emission. 

Our spectral analysis suggests the leptonic contribution to the spatially integrated LMC spectrum is at $\sim$50\%. Admittedly, the parameters fit in our spectral analysis are not well constrained, but the estimated value of the LMC star formation rate is instructive. When all LMC data are included, we estimate the LMC star formation rate to be 0.3 M$_{\odot}$ yr$^{-1}$, assuming $\psi_{\rm MW} = 1$ \citep{2010ApJ...710L..11R}. This is within the error on the rate measured by \citet{2009AJ....138.1243H}, $0.4\pm^{0.4}_{0.2}$ M$_{\odot}$ (see their Figure 11), and given the magnetic field estimate performed in the previous section, we conclude that leptons make a nontrivial contribution to the LMC gamma-ray emission.

In the case of 30 Dor, our star formation estimate is $\sim$2 times the rate measured by \citet{2009AJ....138.1243H} (see their Figure 17). If we take the star formation rate to be the measured value of $0.11\pm^{0.04}_{0.02}$ M$_{\odot}$ yr$^{-1}$, the leptonic fluxes estimated by our spectral fit both decrease by 50\%. The total estimated flux of $2.5 \times 10^{-8}$ ph cm$^{-2}$ s$^{-1}$ is maintained given the anticorrelation between the normalizations of the leptonic and hadronic spectra. Taking this into account, we find that cosmic-ray electrons contribute $\sim$45\% of the 30 Dor emission. This will affect the cosmic-ray diffusion length scales reported by \citet{2012ApJ...750..126M}, who assumed the 30 Dor gamma-ray emission was entirely due to hadronic collisions.

We test the effect of the hydrogen density estimates on the predicted leptonic contribution to the gamma-ray flux. The neutral and ionized hydrogen densities affect the electron energy loss rates due to bremsstrahlung and ionization (Equations~\ref{eq:bBremI}--\ref{eq:bIonization}). As an alternative to $(n_{\rm HI}, n_{\rm H\alpha}) = (2.0, 0.0)$ cm$^{-3}$ used in the analysis described above, we use $(n_{\rm HI}, n_{\rm H\alpha}) = (0.06, 0.06)$ cm$^{-3}$. These densities were adopted by \citet{2013ApJ...773..104C} for their model of the inverse Compton gamma-ray spectrum in the Milky Way. Those authors considered that electrons would not spend most of their time in the highest densities in the mid-plane of the Galactic disk but that magnetic fields cause electrons to migrate to regions of lower hydrogen density.

When taking $n_{\rm HI}= n_{\rm H\alpha} = 0.06$ cm$^{-3}$, we find that the bremsstrahlung gamma-ray spectral model peaks at lower energies ($\sim$200 MeV). The best fit to the LMC gamma-ray spectrum using the new bremsstrahlung spectral model predicts a formation rate of 0.07 M$_{\odot}$ yr$^{-1}$, which is inconsistent with the rate of $0.4\pm^{0.4}_{0.2}$ M$_{\odot}$ measured by \citet{2009AJ....138.1243H}. However, if we fix the star formation rate parameter to 0.3 M$_{\odot}$ yr$^{-1}$ as determined by our previous fit , we find that the leptonic contribution to the LMC gamma-ray flux is still 30\%.

Our spatial analysis also suggests a strong leptonic component to the LMC gamma-ray flux. We found from our gas + radiation models that inverse Compton can explain up to $\sim$50\% of the LMC gamma rays. Admittedly, this is an overestimate given that electron calorimetry, which was not considered for the spatial maps, would require a corresponding bremsstrahlung component $>$50\%. This imperfection aside, both our analysis methods agree that the the leptons contribute nontrivially to the LMC signal detected by {\it Fermi}.

\subsection{Diffusion of Cosmic-rays in the LMC} \label{sec:Diffusion}
We have convolved the star formation rate maps shown in Figure~\ref{fig:HZSF} with the smoothing kernels used by \citet{2012ApJ...750..126M}. We have the assembled a spatial model that includes the three gamma-ray emitting cosmic ray collision processes. This map takes the form:
\begin{eqnarray}
\label{eq:diffModel}
M(x, y) = &\frac{F_{\rm \pi^0}}{F_{\rm tot}}& \sqrt{I_{\rm H\alpha}} (a_1 \psi_{\rm 6.3} + a_2 \psi_{\rm 12.5}) \ast \kappa(l_{\rm CRp}) + \nonumber \\
&\frac{F_{\rm brem}}{F_{\rm tot}}& \sqrt{I_{\rm H\alpha}} (a_1 \psi_{\rm 6.3} + a_2 \psi_{\rm 12.5}) \ast \kappa(l_{\rm CRe}) + \nonumber \\ 
&\frac{F_{\rm IC}}{F_{\rm tot}}& I_{\rm 160 \mu m} (b_1 \psi_{\rm 6.3} + b_2 \psi_{\rm 12.5}) \ast \kappa(l_{\rm CRe}), \nonumber \\
\end{eqnarray}
where the fluxes from each component are taken from the first row of Table~\ref{tab:specFits}, the $a_i$ and $b_i$ values are those given in Tables~\ref{tab:Hadron} and \ref{tab:Lepton} for their respective target model, and $\ast$ is the convolution operator. \citet{2012ApJ...750..126M} use two different smoothing kernels, $\kappa$: the first is a two-dimensional gaussian, $\kappa(l) = e^{-{\bf r}^2 / l^2}$, which models diffusion; the second is an exponential profile, $\kappa(l) = e^{-{\bf r} / l}$, which represents diffusion modified by cosmic-ray energy loss and escape.

\begin{figure}[!ht]
\centering
\includegraphics[width=3in]{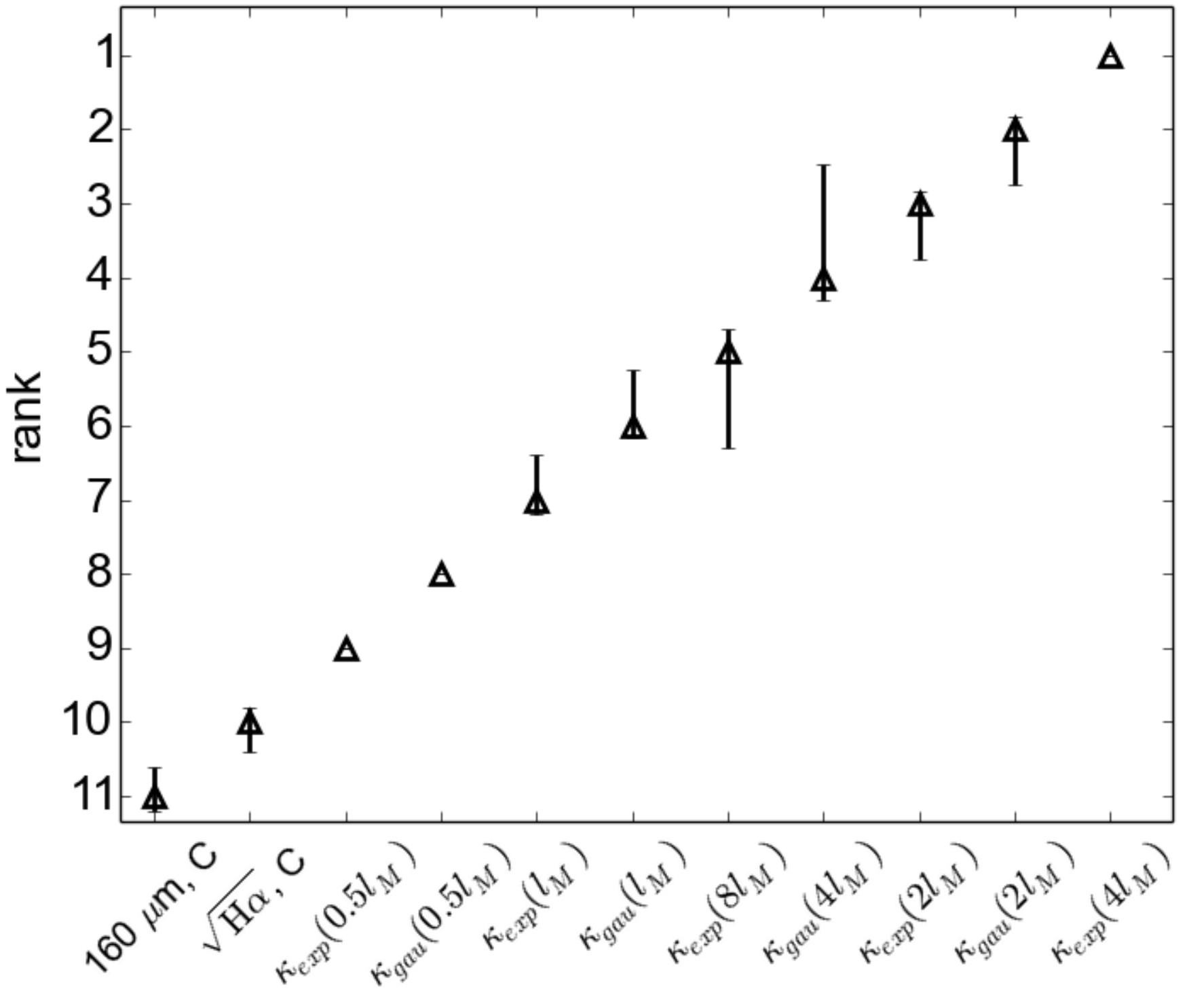}
\figcaption{Rank of diffusion models as determined by the likelihood statistic. Triangles indicate rankings based on the original dataset. Error bars indicate $\pm 1\sigma$ spread on mean rankings determined using ten bootstrapped samples. $l_M$ indicates the best fit length scale found by \citet{2012ApJ...750..126M}, and $\kappa_{\rm exp}$ and $\kappa_{\rm gau}$ refer to the exponential and Gaussian smoothing kernels respectively. \label{fig:diff_rank}}
\end{figure}

In their study, \citet{2012ApJ...750..126M} found the best fit scale lengths for cosmic-ray electrons to be 0.1 and 0.2 kpc for the exponential and gaussian kernels respectively. The scale lengths for protons were found to be 0.2 and 0.45 kpc. Keeping the ratio $l_{\rm CRp} / l_{\rm CRe}$ fixed to the value determined by \citet{2012ApJ...750..126M}, we vary the length scales by factors of two, apply them to the model described by Equation~(\ref{eq:diffModel}), and run each model through the binned likelihood analysis discussed above. Figure~\ref{fig:diff_rank} shows the rankings of various models using diffusion smoothing.

All models that incorporate smoothing kernels outrank their unsmoothed counterparts. We find that doubling and quadrupling the scaling lengths reported by \citet{2012ApJ...750..126M} yield the highest ranked models. The exponential kernel with $(l_{\rm CRp},\ l_{\rm CRe}) = (0.8, 0.4)$ kpc is consistently ranked highest among both the original data set and the bootstrapped samples. This is at least in part due to the fact that \citet{2012ApJ...750..126M} find $l_{\rm CRe} \approx 0.5 l_{\rm CRp}$. Those authors determined only $l_{\rm CRp}$ from gamma-ray data, but we are using both $l_{\rm CRp}$ and $l_{\rm CRe}$ to model the emission measured by {\it Fermi}.

We note that this is a rather crude estimate of diffusion across the entire LMC. Spatial variation of the diffusion scale lengths can be caused by the magnetic structure of the LMC and may play an important role in our results; however, our current analysis cannot address this issue. The complexity of cosmic-ray propagation on the scale of an entire galaxy would be best studied using kinematic simulations such as GALPROP or ideally hydrodynamical simulations \citetext{e.g.\ \citealp{2013ApJ...777L..16B, 2014MNRAS.437.3312S}}, which is beyond the scope of this article.

\subsection{30 Dor as a Point Source and Additional LMC Model Parameters}
The previous analyses of the LMC {\it Fermi} data have been careful to consider the possibility that the gamma-ray emission from 30 Dor may not originate from cosmic-ray interactions, but rather from objects known to produce point-like emission in the gamma-ray sky. Specifically, the presence of pulsars and an X-ray background source are the causes for concern. Recently, HESS has observed one of these pulsars, PSR J0537-6910 and its associated wind nebula N 157B, at TeV energies \citep{2015arXiv150106578H}. The spectrum they fit to the N 157B emission (their Figure 4) suggest a flux of about $2 \times 10^{-11}$ erg cm$^{-2}$ s$^{-1}$ at 20 GeV, which is comparable to our 30 Dor flux estimate at the same wavelength (see Figure~\ref{fig:SpecAnalysis}). However, at 4 GeV, their spectrum suggests the flux from N 157B should be an order of magnitude below the 30 Dor flux measured by {\it Fermi}. This pulsar cannot account for the bulk of the 30 Dor gamma-ray emission at {\it Fermi} wavelengths.

In a talk presented at the Fifth International {\it Fermi} Symposium\footnote{\url{http://fermi.gsfc.nasa.gov/science/mtgs/symposia/2014/program/05_Martin.pdf}}, Pierrick Martin presented a spatial model used by the {\it Fermi} group for their recent analysis of gamma-ray emission from the LMC (publication in preparation, we thank our referee for pointing us to these slides). Their model includes two point sources that lie outside the 30 Dor region. The first is supernova remnant N 132D, the type of object that has motivated our spatial maps because we expect it to be accelerating cosmic rays. The location of N 132D already corresponds to a region of heightened star formation during the 12.5 Myr epoch (Figure~\ref{fig:HZSF}). The second source also coincides with a region of enhanced star formation north of 30 Dor, but has not been positively identified as a supernova remnant. Based on the location indicated on slide 5 of the talk referenced above, we have tested the effect of modeling this additional point source at the position of 2XMM J053444.6-673856. We find the rankings of our gas models remain unchanged, but the LMC flux estimates decrease by $\sim$8\%. For readers familiar with the {\it Fermi} test statistic, the TS for this source is $\sim$90, i.e., the log-likelihood increases by $\sim$45 when we add this point source to the model.

We have also tested the effect of adding an additional spectral parameter to our gas templates. The extra parameter is defined such that the spectrum takes the shape of a parabola in log-space, and Equation~(\ref{eq:fitFunc}) takes the form
\begin{equation}
\label{eq:logParabola}
I_{\gamma}(x, y, E) = \sum\limits_i a_i M_i(x, y) E^{-b_i - c_i \ln(E)}.
\end{equation}
As was the case for the additional point source, we find the rankings of the gas models remain unchanged by the addition of the $c_i$ parameter. The log-likelihoods increase by $\sim$20 (TS$ \sim 40$) and the flux estimates are unchanged within error.

\section{CONCLUSIONS}\label{sec:Conclusions}
The LMC is the brightest and most spatially extended source of diffuse gamma-ray emission outside the Milky Way. It provides a unique laboratory for the study of cosmic-ray physics on global scales. The {\it Fermi} LMC observations are thus complementary with the Milky Way diffuse emission, where smaller structures and even individual supernova remnants can be resolved but where distances to sites of cosmic-ray acceleration may not be well constrained.

We have performed both spatial and spectral analyses of the 5.5 years gamma-ray data from the LMC collected by the {\it Fermi Gamma-ray Space Telescope}. Our spatial analysis proceeded by modeling the gamma-ray data with observed maps of the LMC radiation field and ISM components that are expected to contribute to the gamma-ray emission, namely, star formation and non-thermal radio emission (cosmic-ray flux), ionized, neutral, and molecular hydrogen (target protons for $\pi^0$ decay and bremsstrahlung channels), and dust and infrared stellar emission (target photons for inverse Compton channel). We quantitatively distinguish between these models by ranking them by their likelihood statistic computed by the binned maximum likelihood estimator implemented in the {\it Fermi} Science Tools. We have tested the robustness of these rankings by applying bootstrap resampling to the LMC gamma-ray data set.

From our spatial analysis, we conclude that cosmic rays remain relatively concentrated near their sites of acceleration before losing energy via gamma-ray production. We find the gamma-ray spectral index to range between 2.2 and 2.4 across all of our spatial models. In the case where we simultaneously fit gas and radiation models, we find inverse Compton contributes significantly to gamma-ray flux. A model for inverse Compton emission had not been considered by previous studies of the {\it Fermi} LMC data.

We perform our spectral analysis by first choosing a spatial model (see \S~\ref{sec:had+lep} for model details) solely for the purpose of estimating LMC gamma-ray fluxes in six logarithmically spaced energy bins ranging from 200 MeV to 20 GeV. To this gamma-ray spectrum, we fit a spectral model which includes $\pi^0$ decay \citep{2004A&A...413...17P}, inverse Compton scattering \citep{2013ApJ...773..104C}, and bremsstrahlung (formulated in Appendix~\ref{app:bremSpec}) (see Eqs.~\ref{eq:pionBump}, \ref{eq:inverseCompton}, and \ref{eq:bremsstrahlung}). Our spectral model fit uses least squares, which incorporates errors in the fluxes, to estimate the model parameters.

From our spectral analysis, we find that the cosmic-ray proton power-law spectral index is $\sim$2.4 to within 13\% error. This value is consistent across all three sets of gamma-ray data, full LMC, 30 Dor removed, and 30 Dor isolated. The consistency leads us to conclude that variation in the shape of the cosmic-ray spectrum is not the cause of the spatial variation in gamma-ray flux across the the LMC. Our spectral fits suggests that leptonic processes contribute a significant fraction of the LMC gamma-ray flux, which has not been considered by previous studies. Assuming cosmic-ray energy equipartition with magnetic fields, we compute an LMC magnetic field strength of 2.8 $\mu$G, which is in good agreement with the measurement made by \citet{2005Sci...307.1610G}. 

We have applied the smoothing kernels used by \citet{2012ApJ...750..126M} to the star formation rate maps in Figure~\ref{fig:HZSF} to model cosmic-ray diffusion in the LMC. We find that doubling and quadrupling the scale lengths reported by the \citet{2012ApJ...750..126M} result in the highest ranking models. We expect this is partly due to the fact that those authors determined only the cosmic-ray proton scale length from the {\it Fermi} data, whereas we have fit the data using models for both electron and proton diffusion.

We thank the anonymous referee for suggestions that vastly improved our paper. We thank Sui Ann Mao and Robert Brunner for a useful discussions and Keith Bechtol for helpful recommendations about performing data reduction and analysis using the {\it Fermi} Science Tools software. This work is supported by NASA under the {\it Fermi} Guest Investigator Program (NNX11AO18G and NNX12AO84G) and the University of Illinois Computational Science \& Engineering Fellowship.

\begin{appendix}

\section{NON-THERMAL RADIO MAP}\label{app:nthRadio}
To distinguish the thermal and non-thermal components of the
1.4\,GHz radio emission in the LMC, we use the method presented by
\citet{tabatabaeietal07}. In essence, this method constructs a model
of the thermal radio emission from an extinction-corrected image of a
galaxy's \ha\ emission. The extinction correction is derived by
estimating the dust optical depth from the 160$\mu$m flux density and
the dust temperature. The method is described in detail in Sections 3--6
of \citet{tabatabaeietal07}.  Here we summarize the key
equations, and discuss the values of specific parameters that we have
adopted for the decomposition of the 1.4\,GHz LMC mosaic.

To estimate the dust temperature, we use the map constructed by
\citet{bernardetal08}. From this map, we obtain a map of the
dust optical depth at 160\,\m\ ($\tau_{160}$) according to:
\begin{equation}
I_{160} = B_{160}(T)\left [ 1 - \exp(\tau_{160}) \right],
\label{eqn:tau160}
\end{equation}
where $I_{160}$ is the flux density at 160\,\m, and $B_{160}(T)$ is
the value of the Planck function at 160\,\m\ for dust temperature
$T$. The dust optical depth at 160\,\m\ is typically small, reaching a
maximum value of $\tau_{160} \sim 2.5 \times 10^{-3}$ for several
locations within the chain of molecular clouds south of 30~Doradus.

\citet{tabatabaeietal07} use the extinction curve of the
standard model for dust in the diffuse ISM to convert between
$\tau_{160}$ and the dust optical depth at the wavelength of \ha\
emission $\tau_{\ha}$, adopting $\tau_{\ha} \sim 2200 \tau_{160}$. Since
the large-grain populations in the LMC and Milky Way are likely to have
similar optical properties \citep[compare, for example, Figures 7 and
8 of][]{pei92}, we use the same conversion factor here. If the sources
of \ha\ emission were located behind the galaxy, then the observed
\ha\ emission $I_{\ha}$ would be related to the intrinsic
(i.e. extinction-free) emission $I_{\ha,0}$ via
\begin{equation}
I_{\ha} = I_{\ha,0} \exp(-\tau_{\ha})\;.
\label{eqn:tauha}
\end{equation}
\noindent We use the SHASSA map described in
Section~\ref{sec:Data} to estimate $I_{\ha}$ at each map
pixel. In general, \ha\ sources lie within the galaxy, so $\tau_{\ha}$
provides only an upper limit to the attenuation. The effective optical
depth is $\tau_{\rm eff} = f_{\rm d} \times \tau_{\ha}$, where $f_{\rm
  d} \in [0,1]$ is the dust-screening factor that represents the
relative geometry of the \ha\ emission and the dust that contributes
to the extinction. If the dust, \ha\ sources, and DIG
are well mixed, $f_{\rm d} = 0.5$. For the Milky Way,
\citet{dickinsonetal03} find $f_{\rm d} = 0.33\pm0.15$, indicating
that \ha\ emission has a smaller vertical scale height than the
dust. The porous appearance of the \hi\ emission in the LMC suggests
that the ISM transparency in the LMC may be greater than in the
Milky Way. We adopted  $f_{\rm d} = 0.1$, which
corresponds to a mean extinction in the LMC of $A_{\ha} = 0.2$\,mag.

Having obtained an estimate for the intrinsic \ha\ flux
density, $I_{\ha} = I_{\ha,0} \exp(-\tau_{\rm eff})$, we use
equation~9 of \citet{valls-gabaud98} to estimate the emission measure
EM:
\begin{equation}
I_{\ha,0} = 9.41 \times 10^{-8} T_{e4}^{-1.017} 10^{-\frac{0.029}{T_{e4}}}EM.
\label{eqn:em}
\end{equation}
\noindent In this equation, $T_{e4}$ is the electron temperature
$T_{e}$ in units of $10^{4}$\,K, and the expression is determined
assuming Case B recombination (i.e., each Lyman line photon is
resonantly scattered many times). For twelve \hii\ regions in the LMC,
\citet{vermeijvanderhulst02} derived a mean electron temperature of
10,000\,K. Individual measurements varied between 8000 and 16,000\,K,
depending on the location of the \hii\ region and emission line that
they used in their analysis. The electron temperature of the DIG
in the LMC is not well determined; typical estimates
in the Milky Way range between 8000 and 10,000\,K
\citep[e.g.,][]{reynolds85,alvesetal10}. For our LMC decomposition, we
adopted $T_{e} = 8000$\,K.

\begin{figure}[!ht]
\begin{centering}
\includegraphics[width=3.4in]{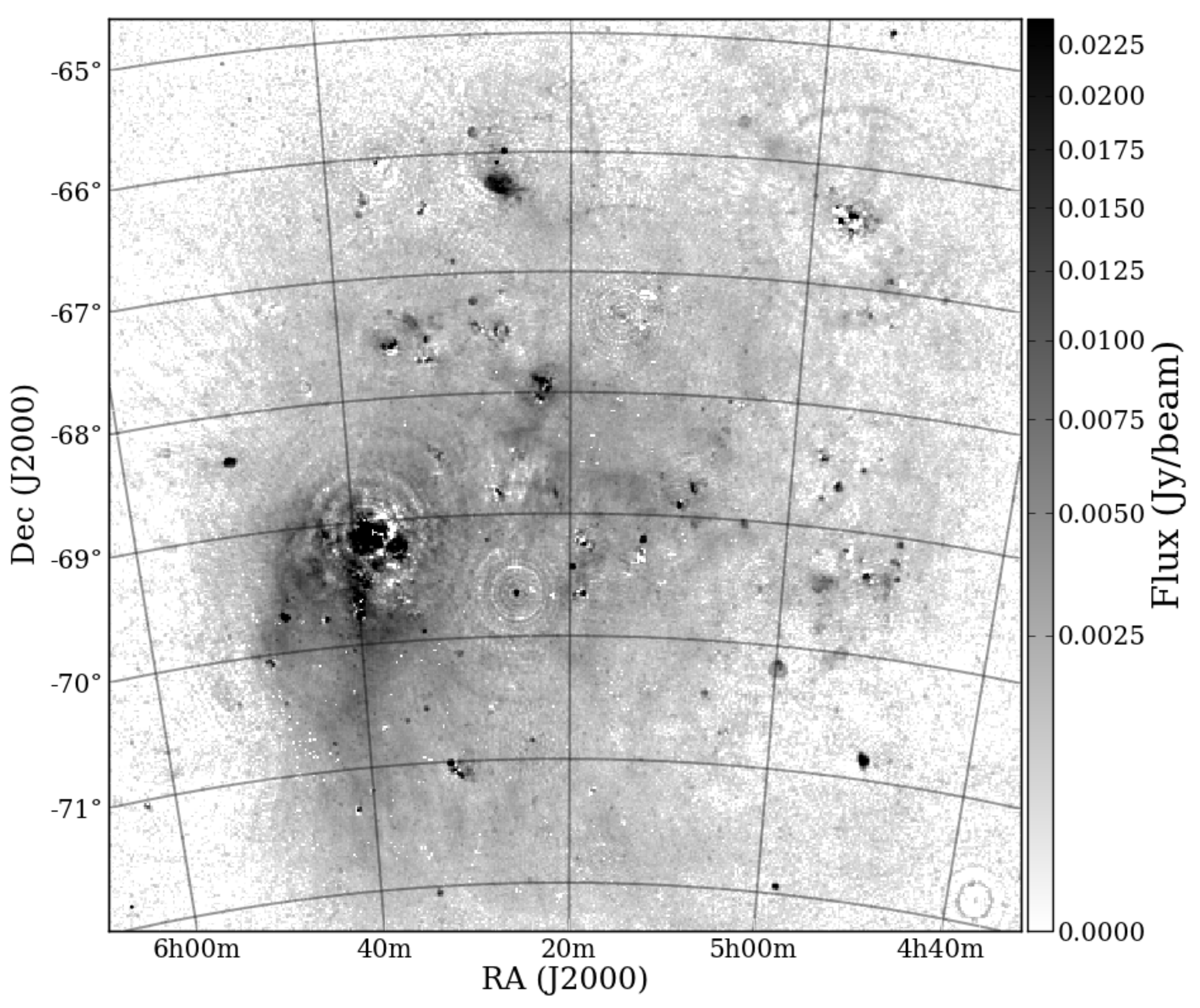}
\figcaption{Non-thermal 1.4 GHz map of the LMC used for our \textit{diffuse} cosmic-ray electron models. See \citet{2007MNRAS.382..543H} for details about the original radio data. \label{fig:nthRadio_full_res}}
\end{centering}
\end{figure}

The optical depth of the radio continuum emission $\tau_{c}$
at frequency $\nu$ is related to the emission measure derived from
Equation~\ref{eqn:em} by
\begin{equation}
\tau_{c} = 8.235 \times 10^{-2} a T_{e} \nu_{\rm GHz}^{-2.1}(1+0.08)EM,
\label{eqn:tauc}
\end{equation}
where $\nu_{\rm GHz}$ is the observed frequency expressed in GHz, and
$a$ is a correction factor that is approximately equal to unity at
1.4\,GHz \citep[see Table~3 of][]{dickinsonetal03}. The predicted
brightness temperature of the free-free radio continuum emission
$T_{\rm b}$ is then simply
\begin{equation}
T_{\rm b} = T_{e}\left ( 1 - \exp(\tau_{c}) \right).
\label{eqn:rctb}
\end{equation}
\noindent To obtain a map of the non-thermal radio continuum emission
at 1.4\,GHz, we subtract this model of the thermal radio emission from
the median-filtered 1.4\,GHz continuum map. The resulting maps of the
LMC's non-thermal 1.4\,GHz radio continuum emission are
presented in Figure~\ref{fig:nthRadio_full_res}. The integrated 
non-thermal flux density is 265\,Jy,
corresponding to a global thermal fraction of 27\%.

The non-thermal map in Figure~\ref{fig:nthRadio_full_res} exhibits a mixture of
diffuse emission and high-brightness features. On one hand, this finding could
indicate that our thermal/non-thermal decomposition is failing for
\hii\ regions, i.e. that we are underestimating the \ha\ extinction
and incorrectly identifying some of the bright thermal 1.4\,GHz
emission as being of non-thermal origin. On the other hand, some
spatial coupling between the brightest sources of thermal and
non-thermal 1.4\,GHz emission would be expected since the massive
stars that ionize the hydrogen gas in \hii\ regions should evolve
quickly to become supernova remnants. Overall, the brightest sources
of non-thermal radio emission demonstrate a good correspondence to the
positions of confirmed LMC supernova remnants \citep{badenesetal10},
which should be strong synchrotron emitters. We find that
$\lesssim20$\% of the LMC's non-thermal emission arises from regions
where the non-thermal radio flux density is greater than
0.2\,mJy\,beam$^{-1}$. A similar fraction is obtained if we calculate
the non-thermal flux density above an \ha\ surface brightness
threshold of 100\,Rayleigh. The majority of the non-thermal radio emission in
the LMC therefore arises from a diffuse, low surface brightness
component that is not directly connected to \hii\ regions.

The method that we have used to separate the thermal and
non-thermal components of the radio continuum emission has a number of
limitations. In particular, we assume that the values of the
dust-screening factor and the electron temperature across the LMC are
constant even though both quantities should vary with interstellar
environment. As noted above, $T_{e}$ may be systematically lower in
the \hii\ regions than in the DIG. There is also empirical evidence
for $T_{e}$ fluctuations within individual \hii\ regions
\citep[e.g.][]{tsamisetal04}. The dust screening factor is probably
not uniform either: the dust distribution may be more clumpy in
star-forming regions than in the diffuse ISM for example. For a
constant dust mass, a clumpy distribution is more porous than a
uniform layer, so the effective \ha\ attenuation will be higher in
regions where the dust and gas are better mixed. A
more sophisticated model would allow $f_{\rm d}$ and $T_{e}$ to vary
across the LMC, but such a model would be difficult to justify without
further independent constraints on $f_{\rm d}$ and $T_{e}$.

\section{BREMSSTRAHLUNG SPECTRUM}\label{app:bremSpec}
To derive Equation~(\ref{eq:bremsstrahlung}), we begin with the gamma-ray emissivity, which can be shown to equal
\begin{eqnarray} \label{eq:qBrem}
q_{\rm brem}(E_{\gamma}) = \frac{dN_{\gamma}}{dV\ dt\ dE_{\gamma}} &=& \frac{4 \alpha}{\pi} \sigma_{\rm T} \ln(183) n_{\rm H} \int_{E_{\gamma}} dE' \frac{\phi_e(E')}{E_{\gamma}} \nonumber \\
&=& \frac{4 \alpha}{\pi} \sigma_{\rm T} \ln(183) n_{\rm H} \frac{\phi_e(>E_{\gamma})}{E_{\gamma}},
\end{eqnarray}
from Equations~(8) to (36) of \citet{1971NASSP.249.....S}. We have taken the ultrarelativistic limit $E_e \gg \alpha^{-1} m_e c^2$ \citetext{Equation~8-28, \citealp{1971NASSP.249.....S}}, and we have assumed the interstellar material is dominated by hydrogen, i.e., $Z \approx 1$. In Equation~\ref{eq:qBrem}, $\phi_e$ is the cosmic-ray electron flux spectrum, $\alpha$ is the fine structure constant, and $\sigma_{\rm T}$ is the Thomson scattering cross section.

We adopt the cosmic-ray electron spectrum and the one-zone model used by \citet{2013ApJ...773..104C} so that we may consistently combine our bremsstrahlung gamma-ray spectrum with their inverse Compton spectrum. The cosmic-ray electron injection spectrum is given by
\begin{equation} \label{eq:injectSpec1}
q_e(E_e) = \frac{dN_e}{dV\ dt\ dE_e} = R_{\rm SN} {\cal N}_e(E_e) V^{-1},
\end{equation}
where $R_{\rm SN}$ is the supernova rate, which is proportional to the star formation rate $\psi$, ${\cal N}_e$ is the number of electrons per unit energy accelerated by each supernova event, and $V$ is the galactic volume. ${\cal N}_e$ is taken to be a broken power law following \citet{2010ApJ...722L..58S}
\begin{equation} \label{eq:injectSpec2}
{\cal N}_e(E_e) \propto \left\{
\begin{array}{lr}
E_e^{-1.8}  & E_e < E_b = 4\ {\rm GeV} \\
E_e^{-2.25} & E_e > E_b = 4\ {\rm GeV}
\end{array}
\right.,
\end{equation}
where $E_e$ is the cosmic-ray electron energy and $E_b$ is the power law break energy. We also assume a hard cutoff at $E_e = 2$ TeV. Taking the LMC as an electron calorimeter and the injection spectrum to be in steady state, the cosmic-ray electron flux is given by
\begin{equation}\label{eq:fluxElec}
\phi_e(E_e) = \frac{dN_e}{dA\ dt\ dE_e} \approx \frac{c}{b(E_e)} \int_{E_e} dE'_e q_e(E'_e) = c \frac{q_e(>E_e)}{b(E_e)},
\end{equation}
where we have assumed $E_e \gg m_ec^2$, and $b(E_e)$ incorporates all the energy-loss mechanisms available for cosmic-ray electrons, i.e.,
\begin{equation}
b(E_e) = b_{\rm IC}(E_e) + b_{\rm sync}(E_e) + b_{\rm brem}(E_e) + b_{\rm ion}(E_e).
\end{equation}
For the injection spectrum given by Eqs.~(\ref{eq:injectSpec1}) and (\ref{eq:injectSpec2}),
\begin{equation}\label{eq:qElec}
q_e(>E_e) \propto \left\{
\begin{array}{lc}
1.25 (E_e^{-0.8} - E_b^{-0.8}) + 0.8 (E_b^{-1.25} - E_{\rm max}^{-1.25}) & E_e < E_b \\
0.8 (E_e^{-1.25} - E_{\rm max}^{-1.25}) & E_b < E_e < E_{\rm max}
\end{array}
\right..
\end{equation}
\citet{2013ApJ...773..104C} give the synchrotron energy-loss rate as
\begin{equation}\label{eq:bSync}
b_{\rm sync}(E_e) \approx 3 \times 10^{-10}\ {\rm GeV\ s^{-1}} \left(\frac{B}{1\ {\rm \mu G}}\right)^2 \left(\frac{E_e}{10\ {\rm TeV}}\right)^2,
\end{equation}
where $B$ is the magnetic field strength. In the Thomson limit where electron energies are low enough that the Klein-Neshina correction to the inverse Compton cross section is unimportant,
\begin{equation}
b_{\rm IC}(E_e) = 2.5 \left(\frac{U_{\rm ISRF}}{1.1\ {\rm eV\ cm^{-3}}}\right) \left(\frac{4\ {\rm \mu G}}{B}\right)^{2} b_{\rm sync}(E_e),
\end{equation}
where $U_{\rm ISRF}$ is the energy density of the ISRF. The bremsstrahlung energy loss rate is given as the sum of two components,
\begin{equation}
b_{\rm brem}(E_e) = b_{\rm brem, i}(E_e) + b_{\rm brem, n}(E_e),
\end{equation}
where 
\begin{equation}\label{eq:bBremI}
b_{\rm brem, i}(E_e) = 1.37 \times 10^{-12}\ {\rm GeV\ s^{-1}} \left(\frac{n_{\rm H\alpha}}{\rm 1\ cm^{-3}}\right) \left(\frac{E_e}{\rm 10\ TeV}\right) \left[\ln\left(\frac{E_e}{\rm 10\ TeV}\right) + 17.2\right]
\end{equation}
is the contribution from cosmic-ray electron interactions with ionized hydrogen and
\begin{equation}\label{eq:bBremN}
b_{\rm brem, n}(E_e) = 7.3 \times 10^{-12}\ {\rm GeV\ s^{-1}} \left(\frac{n_{\rm HI}}{\rm 1\ cm^{-3}}\right) \left(\frac{E_e}{\rm 10\ TeV}\right),
\end{equation}
is the contribution from neutral hydrogen. We use the ionization energy loss term given by \citet{1964ocr..book.....G},
\begin{equation}\label{eq:bIonization}
b_{\rm ion}(E_e) = 7.2 \times 10^{-18}\ {\rm GeV\ s^{-1}} \left(\frac{n_{\rm H}}{\rm 1\ cm^{-3}}\right) \left(3 \ln\frac{E_e}{\rm 10\ TeV} + 2\right).
\end{equation}

\subsection{Milky Way Bremsstrahlung}
As did \citet{2013ApJ...773..104C} for the inverse Compton spectrum, we normalize the bremsstrahlung spectrum to that of the plain diffusion model of \citet{2010ApJ...722L..58S} (see solid cyan curve of right panel of Figure 1). The spectrum of \citet{2010ApJ...722L..58S} is given in terms of specific luminosity 
\begin{equation}\label{eq:Lumq}
L_{\rm brem}(E_{\gamma}) = \int q_{\rm brem}(E_e) dV = q_{\rm brem}(E_e) \int dV,
\end{equation}
where the second equality incorporates the one-zone model assumption. Notice here that the volume integral in Equation~(\ref{eq:Lumq}) cancels with the factor of $V^{-1}$ in Equation~(\ref{eq:injectSpec1}), which means the bremsstrahlung luminosity is independent of galactic volume. From Eqs.~(\ref{eq:fluxElec}), (\ref{eq:qElec}), (\ref{eq:qBrem}), and (\ref{eq:Lumq})
\begin{eqnarray} \label{eq:LofE}
L_{\rm brem}(E_{\gamma}) &=& A^{*} \frac{4 \alpha}{\pi} \sigma_{\rm T} \ln(183) n_{\rm H} c E_{\gamma}^{-1} \nonumber \\
&& \times \left\{
\begin{array}{ll}
\int_{E_{\gamma}} \frac{dE_e}{b(E_e)} [1.25 (E_e^{-0.8} - E_b^{-0.8}) + 0.8 (E_b^{-1.25} - E_{\rm max}^{-1.25})] & E_{\gamma} < E_b \\
\int_{E_{\gamma}} \frac{dE_e}{b(E_e)} [0.8 (E_e^{-1.25} - E_{\rm max}^{-1.25})] & E_b < E_{\gamma} < E_{\rm max}
\end{array}
\right., \nonumber \\
&&
\end{eqnarray}
where all energies are expressed in GeV and $A^{*}$ is the constant to be determined by normalizing to the \citet{2010ApJ...722L..58S} bremsstrahlung spectrum and is proportional to the electron-injection rate, which is proportional to the star formation rate. Again, $E_b = 4$ GeV and $E_{\rm max} = 2$ TeV. For the Milky Way energy loss rates, we adopt the same values as \citet{2013ApJ...773..104C}: $U_{\rm ISRF} = 1.1$ eV cm$^{-3}$, $B = 4\ \mu$G, and $n_{\rm HI} = n_{\rm H\alpha} = 0.5 n_{\rm H} = 0.06$ cm$^{-3}$. Based on the \citet{2010ApJ...722L..58S} bremsstrahlung spectrum, we normalize Equation~(\ref{eq:LofE}) such that $E^2_{\gamma} L_{\rm brem}(E_e) = 10^{40}$ GeV$^2$ s$^{-1}$ GeV$^{-1}$ at $E_{\gamma} = 200$ MeV, which results in $A^{*} = 5.93 \times 10^{29}$ s$^{-1}$.

\subsection{LMC Bremsstrahlung}
The relative importances of the electron energy-loss mechanisms are different in the LMC from those in the Milky Way. In particular, bremsstrahlung losses are more important given the higher neutral-hydrogen density $n_{\rm HI} = 2$ cm$^{-3}$ \citep{2003ApJS..148..473K}. Inverse Compton losses are less important given the lower ISRF energy density $U_{\rm ISRF} = 0.57$ eV cm$^{-3}$, calculated from the LMC SED presented by \citet{2010A&A...519A..67I} where we added the CMB contribution using a blackbody spectrum with $T = 2.73$ K. For the magnetic field strength, we've used $B = 4\ \mu$G \citep{2005Sci...307.1610G}. We also assume the ionized-hydrogen density is negligible compared with the HI density. We use these quantities in the energy-loss-rate equations, and we parameterize the normalization of LMC bremsstrahlung spectrum as $\psi_{\rm LMC} / \psi_{\rm MW}$ resulting in Equation~(\ref{eq:bremsstrahlung}).

\end{appendix}

\end{document}